\documentstyle{mn}

\newif\ifAMStwofonts

\def\ueber#1#2{{\setbox0=\hbox{$#1$}%
  \setbox1=\hbox to\wd0{\hss$\scriptscriptstyle #2$\hss}%
  \offinterlineskip
  \vbox{\box1\kern0.4mm\box0}}{}}
\def \b {\vec}
\def \mathrm{\rm}

\def\Omegam{\Omega_{\rm m}}

\def\lsim{\stackrel{<}{{}_\sim}}
\def\gsim{\stackrel{>}{{}_\sim}}

\def\alphap{{\alpha^{\prime}}}
\def\alphapp{{\alpha^{\prime\prime}}}

\def\d{{\mathrm{d}}}

\def\iras{{\it IRAS}~1.2 Jy}
\catcode`@=11
\def\gsim{\mathrel{\mathpalette\@versim>}}
\def\@versim#1#2{\lower0.2ex\vbox{\baselineskip\z@skip\lineskip\z@skip
       \lineskiplimit\z@\ialign{$\m@th#1\hfil##$\crcr#2\crcr\sim\crcr}}}
\catcode`@=12
\catcode`@=11
\def\lsim{\mathrel{\mathpalette\@versim<}}
\def\@versim#1#2{\lower0.2ex\vbox{\baselineskip\z@skip\lineskip\z@skip
       \lineskiplimit\z@\ialign{$\m@th#1\hfil##$\crcr#2\crcr\sim\crcr}}}
\catcode`@=12
\def\etal{{\sl et al.~}}
\def\gal{{N_{\rm gals}\over V}}
\def\km{${\rm km\; s}^{-1}$}
\newcommand\be{\begin{equation}}
\newcommand\ee{\end{equation}}
\newcommand{\ba}{\begin{array}}
\newcommand{\ea}{\end{array}}

\ifoldfss
  \ifCUPmtlplainloaded \else
    \NewTextAlphabet{textbfit} {cmbxti10} {}
    \NewTextAlphabet{textbfss} {cmssbx10} {}
    \NewMathAlphabet{mathbfit} {cmbxti10} {} 
    \NewMathAlphabet{mathbfss} {cmssbx10} {} 
  \fi
  \ifAMStwofonts
    \ifCUPmtlplainloaded \else
      \NewSymbolFont{upmath} {eurm10}
      \NewSymbolFont{AMSa} {msam10}
      \NewMathSymbol{\upi}     {0}{upmath}{19}
      \NewMathSymbol{\umu}     {0}{upmath}{16}
      \NewMathSymbol{\upartial}{0}{upmath}{40}
      \NewMathSymbol{\leqslant}{3}{AMSa}{36}
      \NewMathSymbol{\geqslant}{3}{AMSa}{3E}

      \let\leq=\leqslant 
      \let\geq=\geqslant 
    \fi
  \fi
\fi 

\ifnfssone
  \newmathalphabet{\mathit}
  \addtoversion{normal}{\mathit}{cmr}{m}{it}
  \addtoversion{bold}{\mathit}{cmr}{bx}{it}
  \newmathalphabet{\mathbfit} 
  \addtoversion{normal}{\mathbfit}{cmr}{bx}{it}
  \addtoversion{bold}{\mathbfit}{cmr}{bx}{it}
  \newmathalphabet{\mathbfss} 
  \addtoversion{normal}{\mathbfss}{cmss}{bx}{n}
  \addtoversion{bold}{\mathbfss}{cmss}{bx}{n}
  \ifAMStwofonts
    \ifCUPmtlplainloaded \else
      \UseAMStwoboldmath
      \makeatletter
      \new@mathgroup\upmath@group
      \define@mathgroup\mv@normal\upmath@group{eur}{m}{n}
      \define@mathgroup\mv@bold\upmath@group{eur}{b}{n}
      \edef\UPM{\hexnumber\upmath@group}
      \new@mathgroup\amsa@group
      \define@mathgroup\mv@normal\amsa@group{msa}{m}{n}
      \define@mathgroup\mv@bold\amsa@group{msa}{m}{n}
      \edef\AMSa{\hexnumber\amsa@group}
      \makeatother
      \mathchardef\upi="0\UPM19
      \mathchardef\umu="0\UPM16
      \mathchardef\upartial="0\UPM40
      \mathchardef\leqslant="3\AMSa36
      \mathchardef\geqslant="3\AMSa3E

      \let\leq=\leqslant 
      \let\geq=\geqslant 
    \fi
  \fi
\fi 

\ifnfsstwo
  \DeclareMathAlphabet{\mathbfit}{OT1}{cmr}{bx}{it}
  \SetMathAlphabet\mathbfit{bold}{OT1}{cmr}{bx}{it}
  \DeclareMathAlphabet{\mathbfss}{OT1}{cmss}{bx}{n}
  \SetMathAlphabet\mathbfss{bold}{OT1}{cmss}{bx}{n}
  \ifAMStwofonts
    \ifCUPmtlplainloaded \else
      \DeclareSymbolFont{UPM}{U}{eur}{m}{n}
      \SetSymbolFont{UPM}{bold}{U}{eur}{b}{n}
      \DeclareSymbolFont{AMSa}{U}{msa}{m}{n}
      \DeclareMathSymbol{\upi}{0}{UPM}{"19}
      \DeclareMathSymbol{\umu}{0}{UPM}{"16}
      \DeclareMathSymbol{\upartial}{0}{UPM}{"40}
      \DeclareMathSymbol{\leqslant}{3}{AMSa}{"36}
      \DeclareMathSymbol{\geqslant}{3}{AMSa}{"3E}

      \let\leq=\leqslant 
      \let\geq=\geqslant 
    \fi
  \fi
\fi 

\ifCUPmtlplainloaded \else
  \ifAMStwofonts \else 
    \def\upi{\pi}
    \def\umu{\mu}
    \def\upartial{\partial}
  \fi
\fi

\title{Breaking the degeneracy of cosmological parameters in 
galaxy redshift surveys}
\author[M. Susperregi]
       {Mikel Susperregi\\
Fisika Teorikoaren Saila, Zientzi Fakultatea, 
Euskal Herriko Unibertsitatea, PO Box 644, 48080 Bilbao, Spain\\
Email: wtpsuxxm@ehu.es}
\date{\today}

\pagerange{0--0}
\pubyear{0000}

\begin{document}

\maketitle

\label{firstpage}

\begin{abstract}
The measurement of cosmological parameters is investigated 
in a representation of the least-action method that uses a 
redshift-space dataset to simultaneously constrain 
the real-space fields $\delta$,$\b v$. This method is robust 
in recovering the entire evolution of the matter density 
contrast and peculiar velocities of galaxies in real space 
from current galaxy redshift surveys. The main strength 
of the method is that it permits us to break the degeneracy 
of the parameters $b$ and $\Omegam$ (customarily measured 
in the ratio $\beta\equiv \Omegam^{0.6}/b$ from redshift-space 
distortions), and these are evaluated in the current 
context separately. The procedure provides a simple numerical 
means to extract as much information as possible from a given 
sample, in the simplest linear bias model, before resorting to cosmic 
complementarity to resolve the degeneracy in the measurement of 
$\Omegam$. The same premise applies to more sophisticated choices 
of bias models. We construct a likelihood parameter 
$\lambda(b,\Omegam)$ to evaluate the relative likelihood of different 
values of $b$ and $\Omegam$. The method is applied to the 
\iras~redshift survey with a low-resolution Gaussian 
smoothing length of 1200 \km within a spherical region 
$x_{\rm max} \sim 15,000$ \km and the reconstructed 
velocity field is then compared with POTENT-reconstructed 
velocities from the Mark III radial-velocity dataset within 
a radius $\sim 5000$ \km, which have been suitably prepared to 
account for Malmquist bias and other systematic errors. The 
analysis yields a likelihood for the parameters that is 
overall consistent with $\Omegam\approx 0.3$ and $b\approx 1.1$, thus 
lending support to a non-vanishing cosmological constant 
$\Omega_{\Lambda}\approx 0.7$ in a flat universe. 
\end{abstract}

\begin{keywords}
large-scale structure of universe -- cosmology 
-- galaxies: distances and redshifts 
\end{keywords}

\section{Introduction}

\begin{figure*}
\centering
\begin{picture}(210,210)
\includegraphics{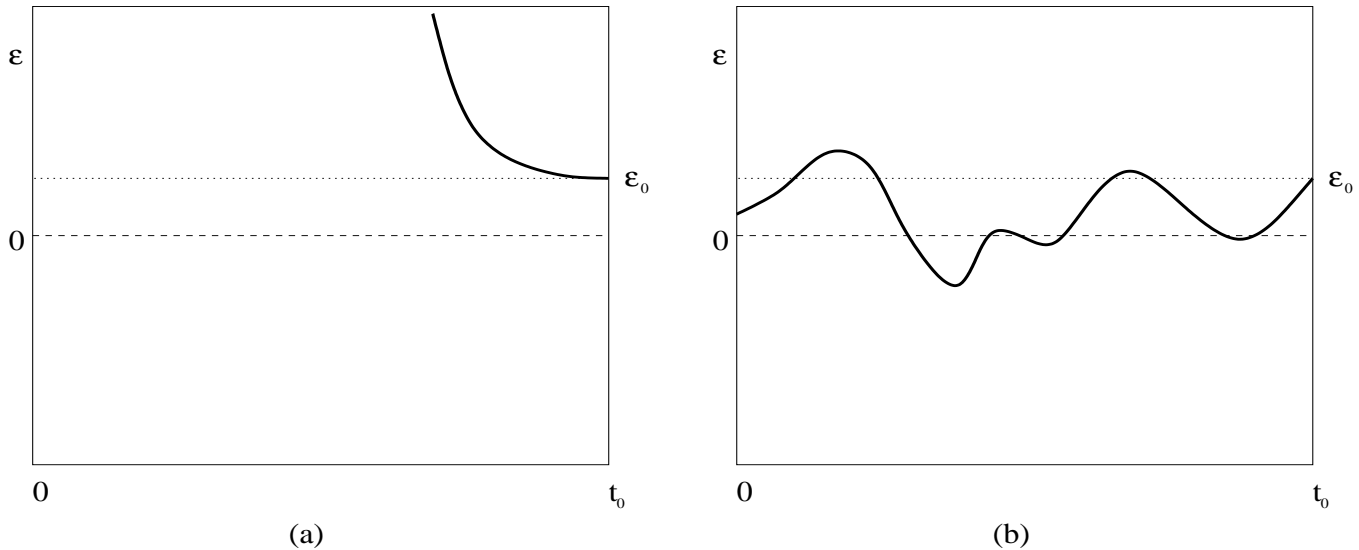}
\end{picture}
\caption[]{Qualitative distribution of errors in the reconstruction 
of the matter fields from redshift surveys. The error $\epsilon_0$ in 
the current redshift sample increases monotonically in (a) as the 
perturbative solutions propagate $\epsilon_0$ to increasing amplitudes 
when integrated back in time; in the boundary condition problem of the 
LAP method, shown in (b), errors fluctuate between the fixed end-points.}
\end{figure*}

Galaxy redshift surveys are undoubtedly extremely valuable tools 
to investigate the evolution of the universe at large scales. 
The cosmologist's prerogative is to determine the evolution 
of the matter density contrast $\delta$ and peculiar 
velocity $\b v$ that yields such cosmic structure, customarily 
assuming that it formed solely by gravity, and the cosmological 
parameters that determine their dynamics. In the standard paradigm 
of a FRW expanding universe, the interplay of both fields is 
governed by the density parameter $\Omega_0$ and the Hubble 
parameter $H_0$. On the other hand, a relationship between 
the fields $\delta$,$\b v$ and the survey data is established 
by adopting a bias model that purports the correlation 
between the $z$-space galaxy number-count and the underlying matter field. 
Devoid of such a relationship, the edifice of measuring cosmological 
parameters from galaxy redshift surveys has no foundation 
whatsoever. A standard working hypothesis, that I shall 
accept throughout this paper, is that of linear bias, 
i.e. $b^2 \equiv P(k)_{\rm gals}/P(k)_{\rm matter}$ 
(more elaborate bias models are propounded in e.g. 
Dekel \& Lahav 1999). For simplicity we shall also 
leave out the scale-dependence of $b$. 
Therefore, three relevant parameters that are interesting 
to pin down from redshift-space samples are in this context 
$\Omegam$,$H_0$ and $b$. In this paper I shall be chiefly 
concerned with $\Omegam$ and $b$ ($H_0$ will be scaled out 
with distance). 

Tracing back in time the matter fields takes us 
to an initial epoch of fluctuations of very small amplitude 
$\delta\lsim 10^{-4}$, seeded by a period of inflationary expansion. 
At that point the information derived from the galaxy 
surveys connects with early-universe data such as 
the spectrum of fluctuations on the CMB. 
If the matter fields could realistically be  
traced back to such a primordial stage by integrating 
the equations of gravitational instability, then 
the statistics of the $\delta$ field would be a potentially key  
discriminant to rule out cosmological models. For instance,
non-gaussianity in the initial $\delta$ field rules out most 
inflationary models, and only those leading to a non-Gaussian 
primordial spectrum remain acceptable (such models are 
suggested in e.g. Linde, Sasaki \& Tanaka 1999). 

Kaiser (1987) proposed measuring cosmological parameters from 
redshift-space distortions by virtue of the fact that overdense 
regions appear to be flatter along the line-of-sight in redshift 
space. This distortion, quantified by the parameter 
$\beta=\Omegam^{0.6}/b$, permits us to solve the equations 
for $\delta$,$\b v$, at least perturbatively (see e.g. Dekel 1994; 
Coles \& Sahni 1995), and measurements of $\beta$ have been
investigated in much detail in the literature (Strauss \& Willick
1995; Dekel 1994,1999a; Dekel, Burstein \& White 1997). 
Also, in view of the fact that the bias parameter is almost 
certainly dependent on the selected sample, estimates have been 
computed for $\beta_{IRAS}$ given $b_I$ for {\it IRAS} galaxies 
(Dekel \etal 1993; Fisher \etal 1995a; Willick \etal 1997a,b; 
Sigad \etal 1998; more recently from the PSC$z$ sample, Canavezes  
\etal 1998; Tadros \etal 1999; Saunders \etal 2000) 
and from the Optical Redshift Survey (ORS) (Hudson \etal 1995; 
Santiago \etal 1995; Baker \etal 1998). The Mark III peculiar 
velocity survey similarly yields estimates of $\beta$ from redshift 
distortions (Willick \etal 1995,1996,1997a,b; 
Dekel, Burstein \& White 1997; Sigad \etal 1998). It is only beyond 
the linear approximation (i.e. $\delta \propto \nabla\cdot\b v$) 
and, indeed, beyond the assumption of linear bias, 
that one can break down the degeneracy between $\Omegam$ and $b$ and 
estimate these parameters separately, rather than via $\beta$ 
(Fry 1994; Bernardeau \etal 1995). Verde \etal (1998) achieved 
this by proposing the bispectrum as a measure of cosmological
parameters, in a model of non-linear bias. In this paper we also 
pursue breaking the degeneracy of $\Omegam$ and $b$ from the
redshift-space data and show that by using the least-action 
framework it is indeed possible to do so within the linear 
bias model. 

The least-action principle (LAP) was first used in 
the Local Group by Peebles (1989,1990). The trajectories 
of nearby galaxies were computed subject to two boundary 
conditions: vanishing initial velocities and fixed 
present positions. This simple scenario of self-gravitating 
point-like masses with two boundary conditions produced an estimate
of $\Omegam$ by fitting to the observations the predicted 
peculiar velocities of nearby galaxies. The LAP method has also been 
used as a test of $\Omegam=1$ CDM models (Branchini \& Carlberg 1994), 
as well as to integrate the orbits of a significant number of galaxies 
from partial coverage redshift samples (e.g. Shaya, Peebles 
\& Tully 1995). An equivalent representation of the LAP method 
in terms of continuous fields, i.e. the density contrast and 
velocity fields was proposed by Giavalisco \etal (1993), 
and employed in Susperregi \& Binney (1994)(hereafter SB94) and 
Susperregi (1995) in the reconstruction of $\Omegam=1$ simple models,  
such as exact solutions and Gaussian random fields. More recently, 
Schmoldt \& Saha (1998) proposed a variant of the customary LAP 
formulation by rewriting the equations motion in redshift space. 

The key difference between the variational and perturbative 
approaches lies on how the errors are spread over the time-reversed 
evolution. This is qualitatively sketched in Fig.~1. A $n$th-order 
solution differs, in the time-reversed direction, from the true solution 
by a monotonically growing parameter $\epsilon$ which sets out from a 
small value $\epsilon(t_0)$ (at any rate $\epsilon_0$ is at 
least the sum of the systematic and random errors of the dataset) 
and the conservation of kinematical quantities is preserved up to 
$O(\epsilon^{n})$. This is adequate within a time span 
$t_c \ll t \lsim t_0$ where $\epsilon(t_c)\sim 1$, and $t_c$ marks 
a transition into the loss of convergence. The distribution 
of errors in the LAP method on the other hand, is by construction 
evenly distributed along the trajectory; the initial and final 
boundary conditions are fixed, though not without systematic 
and numerical errors, and the parameter $\epsilon$ fluctuates 
along the trajectory between both end-points (Fig.~1b). 
Hence the solution is well-behaved whether the errors remain 
within the bound $\epsilon \lsim 1$ or not. In that respect there 
is an advantage with respect to perturbative solutions; the downside 
of it is of course that within the span of time where perturbative 
solutions are valid, LAP errors may fluctuate with larger amplitude 
than the perturbative equivalent. The LAP method, in a nutshell, 
thus consists in finding Ansatze for the matter fields that optimize 
the distribution of $\epsilon$ along the phase-space trajectory, 
and hence minimize the overall departure with respect to the 
exact solution. The following two difficulties may arise: 
\begin{itemize}
\item {\bf A} {\it Finding ``dynamically plausible'' solutions}.  
If the matter field is sparsely sampled or the errors in the dataset 
are substantial, then the boundary condition given by the survey, 
taken at face value, may not correspond to the outcome of 
gravitational evolution from 
the initial fluctuations (typically $\delta\sim 0$ or vanishing peculiar 
velocities). The LAP method will in this case find a {\it dynamically 
plausible} fit between the end-points, which will be as faithful a 
representation of the true evolution as is the quality of the dataset.  
\item {\bf B} {\it Formation of multistreams in over-dense regions}. 
Multistreams are characterised by galaxies at the same redshift which 
are located at different positions along the line of 
sight and have different infalling velocities. The 
degeneracy in redshift among streams makes them indistinguishable 
and hence compatible but inequivalent solutions result, as many as 
there are streams. The LAP method cannot discriminate among 
these solutions; multistreams indeed erase the memory of their 
past evolution.    
\end{itemize}
The second problem can only be overcome by casting aside part of 
the information contained in the sample and smoothing 
over the existing non-linearities to transform the multivalued 
field into a single-valued one, typically with a smoothing length 
$\sim 500 - 1000$ \km. The resulting smoothed field is  
clearly a less resolved representation of the underlying 
galaxy orbits, albeit the only tractable one. 

The advent of large galaxy redshift surveys strengthens the motivation 
to use the LAP method. Near all-sky redshift surveys, e.g. PSC$z$, 
\iras~and ORS provide an excellent sky coverage (within a galactic 
latitude $|b|\gsim 8^{\circ}$ for {\it IRAS} galaxies and 
$|b|\gsim 20^{\circ}$ for ORS), that may be extended further to 
cover the Zone of Avoidance via a Wiener reconstruction  
(Fisher \etal 1995b; Zaroubi \etal 1999). They are therefore 
a fairly thorough representation of the underlying matter 
density field. Obviously the greater number of galaxies 
in the sample the more accurate is the representation 
of the field, and this is best achieved with a redshift survey. 
Real-space datasets require Tully-Fisher distance 
calibrations of individual galaxies, and consequently the end 
result is a sparser sampling than is achieved with the same 
computational effort by measuring redshifts and 
angular coordinates. The goal of this paper is to exploit 
galaxy redshift surveys to the best effect and extract as much 
information from them as is possible; the main thesis put forward is 
the LAP method demonstrably breaks down the degeneracy in the 
determination of $\Omegam$ and $b$. This entails very tangible 
advantages. On the one hand, the freedom to investigate those two 
parameters separately permits us not to take the idea of bias 
seriously. A form of bias will certainly always be present 
in one form or another so that we can make sense of the galaxy 
number-count with respect to the underlying matter field. However, 
whether that is a linear or non-linear bias, the more one 
dissociates this phenomenological relationship from our 
measurements of $\Omegam$, the more credible those measurements 
will be. This is indeed what LAP does. On the other hand, the 
LAP method produces a reconstruction on the basis of the 
redshift-space sample alone, free of any proviso regarding the 
shape of the power spectrum. Assuming a given shape for $P(k)$ 
unduly overconstrains the system, as will the addition of other 
datasets. 

In this article, I shall mainly apply the LAP method to the 
\iras~survey and study the predicted values of $b$ and $\Omegam$. 
The reconstructed \iras~velocity field is then compared with the 
Mark III velocity sample to seek a fine-tuning of the parameters. 
A more thorough undertaking, in terms of the quality of the sample,  
is to apply the LAP method to PSC$z$, which is by a factor 
of 3 a more densely sampled survey than \iras, and it will 
be interesting to tackle this in future work. The article 
is structured as follows: Section 2 describes the LAP method in 
some detail and how to find solutions that are consistent
with a redshift-space dataset; in Section 3 we test the method 
with several {\it IRAS} mock catalogues obtained via $n$-body 
simulations; in Section 4 we apply the method to the \iras~galaxy 
redshift survey, optimizing the predicted velocities with 
the Mark III dataset; finally, in Section 5 we summarize the main 
conclusions.

\section{The LAP method}
\subsection{Redshift-space coordinates}

The redshift coordinates of galaxies are defined  
\be\label{sdef1}
\b z= H_0\b r+\hat r(\hat r\cdot\b v),
\ee
 where $\b r\equiv (r,\theta,\varphi)$ is the physical position, 
$H_0$ is the present value of the Hubble parameter, $\b v$ the 
peculiar velocity, and $\hat r$ a unit vector in the radial 
(line-of-sight) direction. $\b z$ has units of velocity; its 
radial component is the redshift $z_r=cz$, and the angular 
components are the same in both {\it x}-space and {\it z}-space, up to 
the distance scale. Henceforth we shall measure distances in \km,   
hence $H_0$ is scaled out of the equations. In comoving coordinates, 
(\ref{sdef1}) reads
 \be \label{sdef2}
\b s= \b x+ \hat{x}(\hat x\cdot\nabla_{x}\alpha),
\ee
 where the scale factor of the universe is normalized 
to $a(t_0)=1$; $\alpha(t,\b x)$ is the velocity potential, 
$\b v \equiv a^{-1}\nabla\alpha$. Hereafter we adopt $t_0=1$. 

\subsection{Dynamics}

The cosmological perturbations are derived from the action 
\be \label{action}
{\cal S}=\int_0^{1}\!\d t\!\int_{\rm sample}\!\d\b x {\cal L},
\ee
where ${\cal L}$ is given by  
\be  \label{lagrangian}
{\cal L}={1\over 2}(1+\delta){\b v}^{2} +\alpha\xi
-\phi\delta-{|\nabla\phi|^2\over 3\Omegam a^2};
\ee
$\delta$ is the density contrast and $\phi$ the gravitational 
potential caused by the perturbations and 
\be
\xi\equiv \dot\delta+{1\over a}\nabla\cdot[(1+\delta){\b v}]  
\ee
is the {\it excess flux}. The variations 
$\delta{\cal S}/\delta {v_i}=\delta{\cal S}/{\delta\phi}=0$ yield 
\be \label{zveloc-alpha}
{\b v}={1\over a}\nabla\alpha,
\ee
\be \label{zpoisson}
\nabla^2\phi={3\over 2}a^2\Omegam\delta.
\ee
Similarly, 
$\delta{\cal S}/\delta\delta=\delta{\cal S}/\delta\alpha=0$ yield 
respectively 
 \be  \label{zbern}
\xi=0,
\ee
\be \label{zbern2}
\dot\alpha+{|\nabla\alpha|^2\over2a^2}+\phi=0, 
\ee
where we have eliminated $\b v$ via (\ref{zveloc-alpha}) and we 
do not consider $\Omega_{\Lambda}$. The field equations 
(\ref{zbern}),(\ref{zbern2}) are subject to the following 
boundary conditions:

\newcounter{bean}
\begin{list}{\Roman{bean}}{\usecounter{bean}}
\item {\it Homogeneity of the density field at} $t\to0$. Density 
perturbations grow from initial fluctuations of negligible amplitude:
\be \label{bcd}
\delta(t\to0,{\b x})\approx 0.
\ee
\item {\it Galaxy redshift survey at the present time}. The 
galaxy number-count density $\rho_s$ in $z$-space constrains 
the real fields $\delta(\b x)$ and $\alpha(\b x)$ via 
\be \label{constraint}
\rho_s(\b s)=x^2 \gal \biggl({1+b\delta\over 1+\alpha''}\biggr),
\ee
\end{list}  

\noindent where the tilde denotes derivation along the radial 
direction, $x$ is the radial comoving distance and $b$ is the bias 
parameter. Condition $(I)$ is motivated by the CMB Sachs-Wolfe 
constraint $\delta\lsim 10^{-4}$ over $r\sim 100,000$ \km, 
so we accept that perturbations 
are negligible in the limit $t\to 0$. A proof for $(II)$ is given 
in Appendix A. In order to solve (\ref{zbern}),(\ref{zbern2}),  
we construct the trial fields: 
\be \label{zd1}
\delta=\sum_{n=0}^N f_n(t)\delta_n(\b x),
\ee
\be \label{zd2}
\alpha=\sum_{n=0}^N g_n(t)\alpha_n(\b x),
\ee
where the basis functions $f_n$,$g_n$ are adjusted to numerical convenience. SB94  
considered $f_n\equiv D(D-1)^n$, and $g_n=(\dot{D}/D)f_n$, where $D$ is the 
linear growth factor, normalized to unity at $t=1$, so that the lowest-order 
series (\ref{zd1}),(\ref{zd2}) 
are identical to the perturbative solutions. 
This is however strictly speaking not a compelling choice, and a sensible 
choice of orthogonal polynomials leads to an Ansatz of better 
convergence. As we have discussed in the Introduction 
(point {\bf A}), the sparseness of the dataset obscures the 
dynamical evolution and the LAP method is reduced to a numerical fit  
of the fields to the truncated equations, that we derive 
below, subject to (\ref{bcd}),(\ref{constraint}). In trying to 
approximate a function $f(t)$ by orthogonal polynomials $P_m(t)$ 
in $0\lsim t\lsim 1$, a weight function $w(t)\geq 0$ tells us 
the relative importance of the errors spread over the domain. 
For a uniform $w$, $f_n$ are the [spherical] Legendre polynomials 
$L_m(t)$, whereas for a weight function that is larger at the 
endpoints (\ref{bcd}),(\ref{constraint}) than throughout the 
trajectory, e.g. $w(t)=(1-t^2)^{-1/2}$ (by shifting the domain 
from $[0,1]$ to $[-1,1]$), the optimal choice are in this case 
Chebyshev polynomials $T_n(t)$. This choice minimizes the errors 
around the endpoints and it gives a greater weight to the 
solutions (matching the boundary conditions) in this region. 
In the analysis that follows, we shall adopt $f_n=T_n$ and 
$g_n= a^2 f_n$. The fields $\delta_n$,$\alpha_n$ 
are expanded in terms of spherical harmonics,
\be
\delta_n =\sum_{rlm}\delta^{(n)}_{rlm}\,j_l(k_rx)\,Y_{lm},
\ee
\be \label{zalphan}
\alpha_n =\sum_{rlm}\alpha^{(n)}_{rlm}\,j_l(k_rx)\,Y_{lm},
\ee
where $j_l$ are spherical Bessel functions. Substituting 
(\ref{zd1}),(\ref{zd2}) into (\ref{zveloc-alpha}),(\ref{zpoisson}) 
we get
\be \label{v_spher}
\b v = a\!\! \sum_{rlmn} \Big[
\hat{x} \alphap_{rlm}^{(n)}j_l(k_rx) 
+{1\over x}(\hat{x}\wedge\b J_{lm}^{(n)})\Big] T_n Y_{lm},
\ee
\be \label{phi_spher}
\phi=-{3\over 2}a^2\Omegam\!\!\sum_{rlmn}
{\delta_{rlm}^{(n)}\over k_r^2}\,T_n j_l(k_rx)Y_{lm};
\ee 
the coefficients $\alphap_{rlm}^{(n)}$ and $\b J_{lm}^{(n)}$ 
are given in Appendix B. The 
boundary conditions (\ref{bcd}),(\ref{constraint}) then read 
\be \label{constraint1}
0= \sum_{n=0}^N (-1)^n\delta_n,
\ee
\be  \label{constraint2}
\rho_s =x^2 \biggl({N_{\rm gals}\over V}\biggr)
\Big[1+b\,\delta(1,\b x)\Big]
\Big[1+\alphapp(1,\b x)\Big]^{-1},
\ee
where $t$ is rescaled to the interval $[-1,1]$ for convenience 
in using $T_n$, and in (\ref{constraint1}) we have used 
$T_n(-1)=(-1)^n$. The choice of basis functions of SB94 satisfy 
(\ref{constraint1}) by construction, and in our choice of basis 
functions the constraint 
is less trivial, but still it is easily tackled numerically. If we 
restrict ourselves to the interval $0\leq t\leq 1$, then 
(\ref{constraint1}) evaluated at $t=0$ eliminates all the Chebyshev 
polynomials of odd order. This is an 
equivalent approach but we shall adopt the convention above, 
$-1\leq t\leq 1$. The constraint (\ref{constraint2}) is the 
core of the problem as it is where all the information of the dataset 
is contained. The remainder of the paper will focus on the 
different ways one can use that constraint. 

\subsection{Finding LAP solutions}

Substituting (\ref{zd1})--(\ref{zalphan}) into equations 
(\ref{zbern}),(\ref{zbern2}), we get 
\[
\sum_{n=0}^{N}\sum_{rlm}
\Big[\dot{T}_n\delta_{rlm}^{(n)}-k_r^2 T_n \alpha_{rlm}^{(n)}\Big]
\,j_l(k_rx)\,Y_{lm}
\]
\be \label{zceqI}
= -\sum_{p,q=0}^{N}\!\!\!\sum_{\scriptstyle{rlm}\atop\scriptstyle{r'l'm'}}
\!\!T_p T_q\biggl\{{\alphap_{rlm}^{(q)}} 
{{\delta^\prime}_{r'l'm'}^{(p)}}j_l(k_rx) j_{l'}(k_{r'}x)
\ee 
\[
+{1\over x^2}\Big[\hat{x}\wedge\b J_{lm}^{(p)}(\delta)\Big]
\cdot\Big[\hat{x}\wedge\b J_{l'm'}^{(q)}(\alpha)\Big]\biggr\}
Y_{lm}Y_{l'm'},
\]
and
\[
\sum_{n=0}^{N}\sum_{rlm}
\Big[-{3\over 2}\Omegam k_r^{-2}\delta_{rlm}^{(n)}
+\biggl({\dot{T}_n\over T_n}+2{\dot a\over a}\biggr)\alpha_{rlm}^{(n)}\Big]
T_n j_l(k_rx)Y_{lm}
\]
\be \label{zceqII}
= -{1\over 2}\!\!\sum_{p,q=0}^{N}\!\!
\sum_{\scriptstyle{rlm}\atop\scriptstyle{r'l'm'}}\!\!
T_p T_q\biggl\{\alphap_{rlm}^{(q)}\alphap_{r'l'm'}^{(p)}
j_l(k_rx)j_{l'}(k_{r'}x)
\ee
\[
+{1\over x^2}\Big[\hat{x}\wedge\b J_{lm}^{(p)}(\alpha)\Big]
\cdot\Big[\hat{x}\wedge\b J_{l'm'}^{(q)}(\alpha)\Big]\biggr\}
Y_{lm}Y_{l'm'},
\]
where the coefficients 
$\b J_{lm}^{(p)}(\delta)$,$\b J_{lm}^{(q)}(\alpha)$ are defined as in 
(\ref{jlmn}) in Appendix B and ${\delta^\prime}_{rlm}^{(n)}$ 
as in (\ref{alphap}) via the trivial substitution $\alpha\to\delta$. 
By multiplying (\ref{zceqI}),(\ref{zceqII}) by $T_rj_lY_{lm}$
and integrating over all coordinates, we get
\[
\sum_{n=0}^{N} \langle T_r\dot T_n\rangle 
C^{\delta}_y\delta_{y}^{(n)}+ \sum_{n=0}^{N}
\langle T_r T_n\rangle C^{\alpha}_{y}\alpha_{y}^{(n)}
\]
\be\label{zshortI}
= -\sum_{p,q=0}^{N}\langle T_r T_p T_q\rangle\!\! 
\sum_{y'y''}D^{y}_{y'y''}
\delta_{y'}^{(p)}\,\alpha_{y''}^{(q)},
\ee
\[
\sum_{n=0}^{N} \Omegam \langle T_r T_n\rangle  
S^{\delta}_{y}\delta_{y}^{(n)}
+ \sum_{n=0}^{N}
\langle T_r (\dot{T}_n+2{\dot a\over a}T_n)\rangle 
S^{\alpha}_{y}\alpha_{y}^{(n)}
\]
\be\label{zshortII}
= -\sum_{p,q=0}^{N}\langle T_r T_p T_q\rangle \!\!
\sum_{y'y''} E^{y}_{y'y''}
\alpha_{y'}^{(p)}\,\alpha_{y''}^{(q)},
\ee
where $y\equiv (rlm)$ and the angle brackets $\langle\rangle$ 
for the Chebyshev polynomials are defined in Appendix C. In deriving 
(\ref{zshortI}),(\ref{zshortII}), the coefficients $C^{\delta}_{y}$, 
$C^{\alpha}_{y}$, $S^{\delta}_{y}$, $S^{\alpha}_{y}$, 
$D^{y}_{y'y''}$ and $E^{y}_{y'y''}$ are calculated via Clebsch-Gordan 
coefficients for cross-products of $Y_{lm}$ and via the standard 
orthogonality relations for $Y_{lm}$ and $j_l$, given in Appendix D.  
Cross-products of $j_l$ terms are estimated numerically. 
We proceed to solve (\ref{zshortI}),(\ref{zshortII}) numerically 
with the following iterative procedure. We first construct an 
Ansatz of the coefficients ${\delta}_{y}^{(n)}$,${\alpha}_{y}^{(n)}$ 
that satisfies, to linear order, (\ref{zshortI}),(\ref{zshortII}) 
as well as (\ref{constraint1}),(\ref{constraint2}). We start out with 
the galaxy number-count density $\rho_s$. Following its definition in 
Appendix A, this quantity has units of inverse velocity, and we 
define its associated $z$-space density contrast via
\be \label{delta_s}
\rho_s \equiv {4\pi N_{\rm gals}\over s_{\rm max}} 
( 1+\delta_s),
\ee
where $s_{\rm max}\equiv cz_{\rm max}$ is the maximum redshift 
in the sample. Our first Ansatz entails $b=1$ and linear evolution, 
so that $\delta_s \propto -\nabla^2\alpha$, and on inverting this 
relation to obtain the coefficients $\alpha_y^{(n)}$, we estimate 
$\delta(\b x)\propto \delta_s (\b x+ \hat{x}\alpha^{\prime})$ by using 
the expression for the radial derivatives (\ref{alphap}). This 
yields a first Ansatz for $\delta_y^{(n)}$, $\alpha_y^{(n)}$,  
derived from the dataset, that satisfies the linearized equations,  
given by the LHS of (\ref{zshortI}),(\ref{zshortII}):
\be \label{homog}
\left[\ba{cc}
 {\rm C}^{\alpha} & {\rm C}^{\delta} \\
 {\rm S}^{\alpha} & {\rm S}^{\delta}
 \ea
\right]\left[\ba{c}
{\b\alpha_y}\\
{\b\delta_y}
\ea
\right]\approx 0,
\ee
where the column vectors are $(\b \alpha_y)_r= \alpha_y^{(r)}$ and 
$(\b \delta_y)_r= \delta_y^{(r)}$, with $r=0,\dots, N$. 
The solutions of the homogeneous system are then re-adjusted to 
satisfy (\ref{constraint1}),(\ref{constraint2}) and we use these 
to construct the quadratic terms on the RHS of 
(\ref{zshortI}),(\ref{zshortII}). This leads to an inhomogeneous 
system that again we solve for $\b \delta_y$,$\b \alpha_y$. On 
each iteration we improve the solutions by least-squaring them to satisfy 
(\ref{constraint1}),(\ref{constraint2}) to the best accuracy and 
we are also free to vary the parameters ($b$,$\Omegam$) for improved 
convergence. This procedure is very accurate, as we will show in 
the next sections, and it permits us to improve the estimate of 
the mapping $\b x\to \b s$ at each iteration using the full 
non-linear relationship (\ref{constraint2}). At each iteration, 
the fields $\b \delta_y$,$\b \alpha_y$ are used to obtain an 
estimate $\tilde\rho_s(\b s)$ of the RHS of (\ref{constraint2}). 
We then vary these fields to obtain a minimum of the quantity 
$\sum_{\b s}(\rho_s-\tilde\rho_s)^2$.  Therefore we do not  
perform a $j_lY_{lm}$ expansion of the dataset, and it 
is very convenient not to do so, as a relationship of this 
kind between the redshift and real-space coordinates entails  
that we compare them via a Taylor expansion $j_l(k_rs)\approx 
j_l(k_rx)+k_r\alphap j_l^{\prime}$; an approximation of this kind 
$\sim {\cal O}(\partial^2 j_l)$ introduces an error of up to 15\% 
for $l\gsim 10$ as can be shown from (\ref{jprime}) in Appendix B. 

\subsection{Normal modes}

We have noted that the linearized equations (\ref{homog}) 
are a homogeneous matrix system. If the determinant of the matrix 
is non-zero, then the only possible solution is $\b \delta_y=0$ 
and $\b \alpha_y=0$. We know however that (\ref{homog}) is also valid 
for linear fields, and these have non-vanishing coefficients. 
Therefore we conclude that the determinant of the system vanishes. 
Such a system of equations is tackled through the Singular Value 
Decomposition (SVD) procedure. It factorises the singular matrix 
in (\ref{homog}) in a product of three matrices: two orthogonal 
matrices U and V, and a diagonal one W, which has one 
or more vanishing {\it weights} along the diagonal. After SVD, 
(\ref{homog}) reads
\be \label{SVD}
\mathrm{U}
 \left[\ba{cccc}
   0&&& \\
    &w_1&& \\
    &&w_2& \\
    &&&\ddots 
 \ea
\right]
\mathrm{V}\left[\ba{c}
{\b\alpha_y} \\
{\b\delta_y}
\ea\right]
=0,
\ee
where the weights $w_1,w_2,\ldots w_N$ are non-zero real numbers. Therefore, 
the vector 
\be
\b N_y=\mathrm{V}\left[\ba{c}
{\b\alpha_y} \\
{\b\delta_y}
\ea\right]
\ee
gives a coordinate basis on which the first component, the 
{\it normal mode}, is unconstrained by the system (\ref{homog}). 
$N_y^{(0)}$ is solely determined by 
(\ref{constraint1}),(\ref{constraint2}). The rest of the components 
of $\b N_y$ (which are identically zero for linear fields) are 
functions of the normal mode. Therefore, one can rewrite the full 
non-linear system (\ref{zshortI}),(\ref{zshortII}) in terms of the 
fields $\b N_y$ and this would be strictly speaking the natural 
basis to investigate the underlying mode coupling induced by 
gravity. In the Fourier formulation with a set of basis 
functions like those used in SB94, $f_n=D(D-1)^n$, it is easy 
to show numerically that the $\b k$-th normal mode is 
given by 
\be \label{k-normal}
N_{\b k}^{(0)}= \delta^{(0)}_{\b k}+k^2\alpha^{(0)}_{\b k}. 
\ee 
This has a simple physical interpretation: (\ref{k-normal}) is a
vanishing scalar for linear fields and thus its departure from zero 
gives us a measure of non-linearity. This quantity is determined 
by the boundary conditions. In the spherical harmonic formulation, 
the normal modes (equivalent to (\ref{k-normal})) are 
\be  \label{normal}
N_y^{(0)}= \sum_{n=0}^N h_n (\delta_y^{(n)}-k_r^{-2}\alpha_y^{(n)}),
\ee
where 
\be 
h_n= {\eta\over \pi}\int_0^1 \!\!\d t w(t)D(t) T_n  
\ee 
where $\eta=1$ for $n=0$ and $\eta=2$ otherwise. The quantity 
(\ref{normal}) vanishes in the linear regime and, like
(\ref{k-normal}), its departure from zero is a measure of 
non-linearity.  

\subsection{Using the method in practice}

The apparent mathematical complexity of the LAP method 
has precluded its wider use in practice. The fraction of papers 
in the literature that employ LAP techniques to investigate 
large-scale structure is minute in contrast to analyses 
based on perturbation theory techniques, such as POTENT, VELMOD 
and others. The latter unquestionably have the virtue of simplicity, 
and are as efficient as they are easy to implement. 
However, in practice the method described in this 
section entails no more complexity than programming an $n$-body 
code; an undertaking that merits the effort, so as to estimate 
$b$ and $\Omegam$, rather than merely $\beta$. The chief difficulty 
resides in writing an algorithm for an effective numerical resolution of 
(\ref{zshortI}),(\ref{zshortII}). This may be a somewhat arduous 
task, but at any rate a very straightforward one with a very basic 
grasp of numerical methods. 

The LAP method is very flexible in its implementation. The basic 
input in the problem are the boundary conditions (\ref{bcd}),
(\ref{constraint}) and the procedure that is to be followed 
to find a stationary action linking both end-points is largely 
a matter of numerical convenience. The algorithm used in this section 
employs Chebyshev polynomials to fit the trial fields $\delta$ and 
$\alpha$ to the dynamics. A myriad of other choices (e.g. binomial 
expansions, Legendre and Hermite polynomials, etc) is also feasible 
and thus the LAP implementation set out above is by no means 
a straightjacket recipe (for a more condensed presentation of 
the algorithm, see Susperregi 2000).  

In short, the algorithm can be summarized as follows. 

\begin{itemize}
\item A galaxy redshift survey is a dataset ${\cal D}$ of points 
($z$,$\varphi$,$\theta$). Those raw data are transformed to 
a smoothed redshift-space field $\rho_s(\b s)$, given a smoothing 
length and a window function $W(k)$. In this article we shall 
exclusively implement Gaussian smoothing. 
\item The name of the game is to compute a fit for $\delta$,$\alpha$. 
The starting point is to make a linear Ansatz that is consistent 
with $\delta_s$, which is derived from (\ref{delta_s}). This is 
achieved by inverting the relation $\delta_s\propto -\nabla^2\alpha$ 
and next estimating $\delta\propto \delta_s(\b x
+\hat{x}\alpha^{\prime})$. 
\item The linear Ansatz is the first input to be used in equations 
(\ref{zshortI}),(\ref{zshortII}). These yield the homogeneous system 
(\ref{homog}), which is our second port of call. The solutions 
$\delta_y$,$\alpha_y$ obtained are least-square fitted to 
(\ref{constraint1}),(\ref{constraint2}). This requires adopting 
a value of $b$.  
\item The adjusted values of $\delta_y$,$\alpha_y$ are brought back 
to construct the RHS of (\ref{zshortI}),(\ref{zshortII}), and from 
there one computes the new $\delta_y$,$\alpha_y$ in the LHS of 
(\ref{zshortI}),(\ref{zshortII}). This part of the operation entails 
an assumed value for $\Omegam$. In the normal mode coordinates 
discussed in 2.4, the modes $\delta_y$,$\alpha_y$ of the cosmic 
fields are merely excitation modes of a harmonic oscillator and 
the terms in the RHS of (\ref{zshortI}),(\ref{zshortII}) represent 
nonlinear perturbations of those excitation modes. 
\item Successive iterations of the procedure eventually yield the 
correct values of $\delta_y$,$\alpha_y$. The values of $b$ and 
$\Omegam$ are readjusted in the process and their estimated 
values are those that result in the most rapid 
convergence of the solutions. 
\end{itemize}

The algorithm thus produces the cosmic fields and an estimate 
of the cosmological parameters. In the remainder of the article 
we shall investigate how to make the best use of the procedure 
and how to quantify the relative likelihood of different values 
of the cosmological parameters. 

\section{Test of the method} 

\begin{figure*}
\centering
\begin{picture}(360,360)
\includegraphics{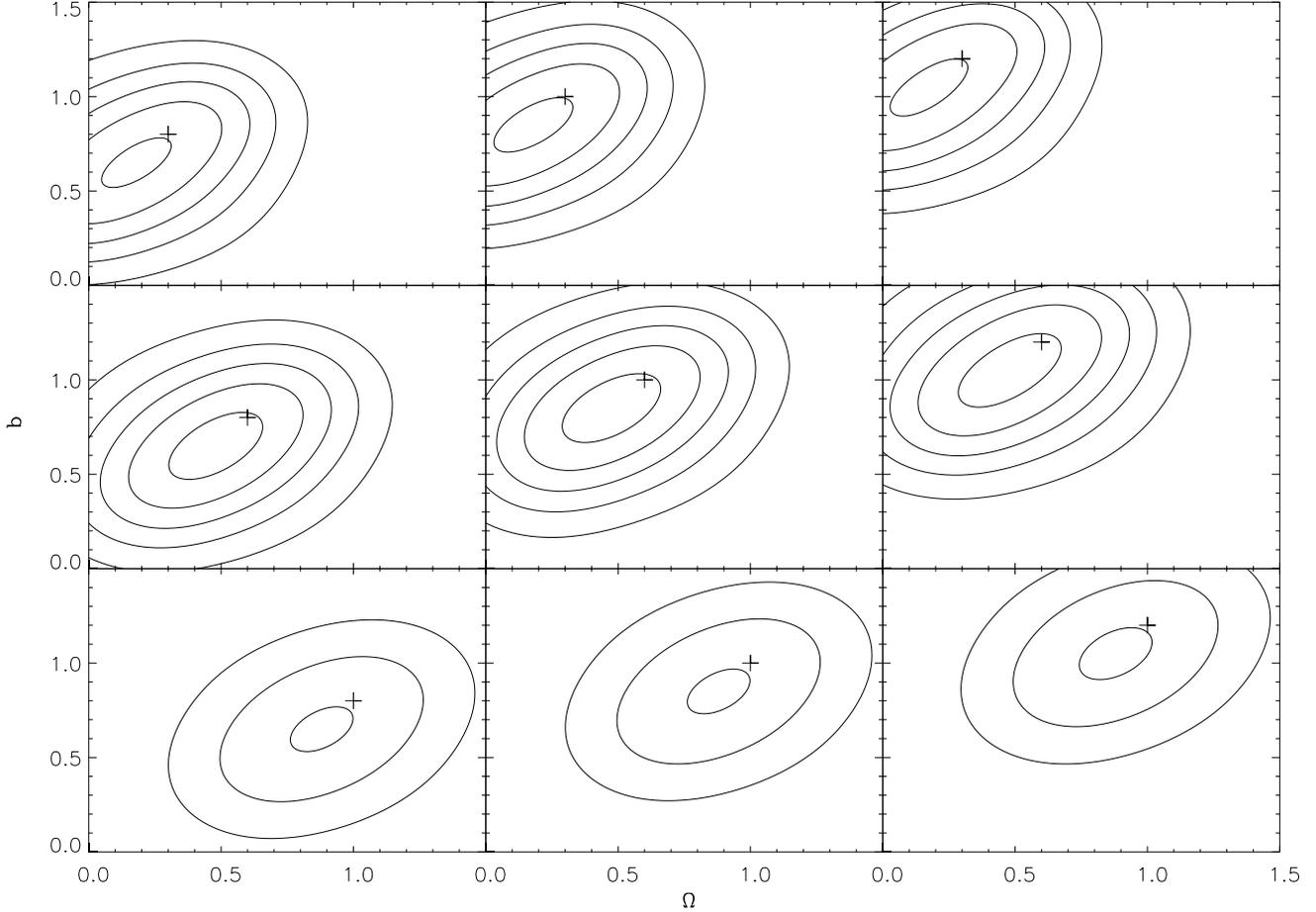}
\end{picture}
\caption[]{Likelihood contours for the reconstruction of the nine 
datasets $d(b,\Omegam)$. The cross on each panel indicates the real
values of $(b,\Omegam)$ in each reconstruction, and the likelihood 
contours are computed following (\ref{likelihood}) with a suitable 
normalization. The concentric contours represent a likelihood of 
95\%, 75\%, 50\%, 25\% and 10\% from the inner curves to the outer, 
on the two upper rows, and 95\%, 75\% and 50\% on the lower row.}
\end{figure*}

We test the LAP method on mock catalogues derived from $n$-body 
simulations, using a Gaussian smoothing length of 600 \km. The 
\iras~power spectrum (Fisher \etal 1993) is adopted as a prior, 
and the simulations are performed over a periodic box 
$L= 25,600$ \km~with $128^3$ grid-points and $128^3$ particles. 
The simulations are performed from Gaussian initial conditions, 
for the following values of the parameters: $b=0.8, 1.0, 1.2$ and 
$\Omegam=0.3, 0.6, 1.0$. The fields are evolved forward in time 
until $\sigma_8\approx 0.7$ over $\sim 800$ \km, using a Gaussian 
cutoff. We choose a two-powerlaw functional form of selection 
function (Yahil \etal 1991): 
\be
\phi(r\geq r_s)=\Big({r_s\over r}\Big)^{2\alpha}\Big(
{r_*^2 + r_s^2\over r_*^2 + r^2}\Big)^{\beta},
\ee
and $\phi(r\leq r_s)=1$, where $r_s= 500$ \km, 
$r_* = 5034$ \km, $\alpha= 0.483$ and 
$\beta=1.79$ (Fisher \etal (1995a); we adopt the estimated 
central values of these parameters and will not test the fine 
detail of the variations of $\phi(r)$ due to their errors),  
and thus we compute the redshift-space dataset over a sphere of radius 
$x_{\rm max}\sim 17,000$ \km. The resulting mock catalogue has 
an effective radius of $\sim 13,000$ \km~beyond which the galaxy 
number-count is sparse and is cut off for the purpose of the
reconstruction. The number of realizations are nine in total, 
and we denote $d(b,\Omegam)$ the $z$-space mock samples 
derived in this way. Each dataset $d(b,\Omegam)$ results 
from a unique pair of real-space fields $\delta$,$\alpha$ 
which are the density contrast and velocity potential that 
we obtain via the $n$-body simulations. 

The tests are carried out by using $d(b,\Omegam)$ as an input dataset in 
(\ref{constraint2}) without any prior assumption on the real values of 
the parameters of the mock sample. We use (\ref{constraint2}) to solve
(\ref{zshortI}),(\ref{zshortII}) following the iterative procedure
given in \S 2.3 and derive the estimated fields 
$\tilde\delta$,$\tilde\alpha$ for different values of the parameters 
$\tilde b$,$\tilde \Omegam$. The likelihood of these parameters is 
estimated on the basis of the performance of the solutions 
$\tilde\delta$,$\tilde\alpha$ in terms of their convergence and 
ability to satisfy the constraints. We use a likelihood 
function given by the inverse of the $\chi$-squared sum of the 
differences of the fields between successive iterations in solving 
(\ref{zshortI}),(\ref{zshortII}), i.e. 
\be \label{likelihood} 
\lambda(b,\Omegam) \propto \Big[\sum_x 
\Big({\delta_n-\delta_{n-1}\over\sigma_{\delta}}\Big)^2
+\Big({\alpha_n-\alpha_{n-1}\over\sigma_{\alpha}}\Big)^2
\Big]^{-1},
\ee
where $n\geq 25$, $\sigma_{\delta}\approx 0.20$, 
$\sigma_{\alpha}$ is a normalization factor that 
rescales the coefficients $\sigma_n$ so that $\alpha$ becomes 
a dimensionless field within the range $-1\lsim \alpha \lsim 1$  
and we have used $N=10$ and $l\leq 15$ and an initial linear Ansatz. 
The results are shown in Fig.~2. The likelihood contours are the 
LAP reconstructions and the crosses on all nine panels of Fig.~2 
indicate the values of the real parameters in each mock
dataset on the $(b,\Omegam)$ plane. As can be observed, 
the likelihood contours are certainly well correlated 
with the location of the crosses, where the innermost contours  
mark the level of 95\% likelihood, that in all cases lie in the 
neighbourhood of the real values of the parameters. The likelihood 
contours show an approximately elliptical shape, with the major 
semiaxes tilted at approximately 45 degrees, suggesting a correlation 
between both parameters that merely arises in the numerical 
computation. The estimates in the reconstruction are fairly good, 
with a trend in underestimating slightly the values of both
parameters. The best reconstructions are for the intermediate 
value of the density parameter $\Omegam=0.6$, shown in the second 
row. In this case the crosses actually lie within the highest
likelihood contours, with very little scatter. Overall, in the 
nine reconstructions the rms scatter in $b$ and $\Omegam$ lie within  
the region $0.26\lsim \sigma^2_{\Omega}\lsim 0.44$, 
$0.15\lsim \sigma^2_{b} \lsim 0.32$. The largest scatter in 
$\Omegam$ is for $\Omegam=0.6$, and a similar situation 
arises with $b$, whereby the intermediate value $b=1.0$ has the 
larger error. 

The effect of underestimating the true values of the parameters 
is systematic and can be calibrated. This effect can be largely  
ascribed to the unconventional choice of likelihood estimator 
(\ref{likelihood}). One could argue that, for slowly varying 
variances, $\lambda\propto b^{-2}$ (chiefly from the $\delta$ 
part of the RHS of (\ref{likelihood})) and therefore smaller values 
of the bias factor (and consequently, by correlation, also of 
$\Omegam$) are favoured.  
However, it is not straightforward to disentangle the 
dependence of the solutions on the parameters after successive 
iterations. The likelihood estimator used is thus to some extent 
biased. However, we find that the criterion of convergence 
given by the RHS of (\ref{likelihood}), suitably normalised, 
is the sharpest discriminator to pin down the best estimates of 
the cosmological parameters. We have carried out numerous tests 
with more conventional likelihood estimators (e.g. Fisher likelihood 
matrix, etc) obtaining much poorer results than with (\ref{likelihood}). 

Fig.~3 shows the density constrast reconstructions for the same 
datasets $d(b,\Omegam)$. The reconstructed density contrast 
$\delta_{\rm LAP}$ is shown on the horizontal axis plotted 
point-by-point within the selected spherical volume ($r\sim 
13,000$ \km) against the real density contrast of the mock datasets. 
A solid line of slope 1.0 is plotted across each panel that 
does not correspond to the regression line on each panel 
though the differences are tiny. The slopes of the regression 
lines lie within the range $0.99\pm 0.08$. The rms value 
corresponding to the random and numerical errors lies in the 
range $0.19\lsim \sigma_{\delta} \lsim 0.28$. The reconstructions 
in Fig.~3 have been carried out with a prior knowledge of 
the values of $b,\Omegam$ for each dataset. Alternatively, 
the test can be carried out by putting together the procedure 
followed to obtain the likelihood in Fig.~2 and investigate 
the scatter resulting in the plots $\delta_{\rm mock}$ vs. 
$\delta_{\rm LAP}$ for different values of $b$,$\Omegam$. 
Supposedly estimating the values of $b,\Omegam$ and finding the 
optimal correlation between $\delta_{\rm mock}$,$\delta_{\rm LAP}$ 
ought to be two not unrelated operations. However these two 
appear to be fairly independent: it turns out that whereas 
(\ref{likelihood}) gives us the correct likelihood estimates 
following the criterion of convergence of the solutions at each 
iteration, the variations in $\sigma_\delta$ for a large range 
of $b$,$\Omegam$ are fairly small, and $\sigma_{\delta}$ (as computed 
from tests such as the nine reconstructions in Fig.~3) is too 
insensitive to be helpful in the estimate of the parameters. 
Therefore the tests show that the estimate of the parameters 
and the reconstruction of the fields are two operations that 
are to a large extent independent. For an arbitrary sample, one 
would thus first compute (\ref{likelihood}), pick the values of 
$b,\Omegam$ at the maximum of the likelihood surface and use 
these to solve the equations to compute $\delta$,$\alpha$. 
Similarly, Fig.~4 shows the comparison of the LAP results with 
the mock data in the reconstruction of the velocity potential. 
The values of the fields have been scaled to $\alpha_{\rm max}$ 
and are dimensionless. It is apparent that the regression line 
is in all cases slightly greater than unity, with a more accentuated 
tilt for larger values of ($b$,$\Omegam$). The smaller values of
$\alpha$ adjust better to a slope of unity, but with larger 
scatter than larger $\alpha$. 

Fig.~5 shows a cross-section on the $Z=0$ plane of a particular 
velocity field reconstruction, that of the dataset $d(b=1.0,\Omegam=0.3)$. 
The figure shows several prominent features of the underlying 
density field in this case: three overdense regions to which 
the field vectors converge, on the lower left, middle right 
and upper left parts of the panel, and two prominent underdense 
regions, from which the velocities diverge, one at the central 
region and another one at the middle-left boundary of the 
circle. It is apparent that the LAP velocities are not vanishing 
in the normal direction of the boundary surface of the selected 
subvolume, and therefore the customary Neumann spatial boundary 
conditions employed on spherical Bessel functions (i.e. vanishing 
normal velocities at the boundary) do not apply. We note that 
spatial boundary conditions are unnecessary in the LAP
reconstruction, thus we have not brought up the issue in \S 2. 
The velocity field agrees within 10\% accuracy with the $n$-body 
exact field within 78\% of the selected volume, and the remaining 
22\% differs from the mock sample velocities by an error of 
$\gsim 10\%$ (shown in Fig.~4 by the regions enclosed by the 
solid curves) and withing this volume 6\% differs by an error 
$\gsim 20\%$ (regions enclosed by broken curves). These regions 
are mostly located in the neighbourhood of peaks, right at the very 
slopes, where the largest velocities are found. The central regions 
of peaks and troughs are very accurately reconstructed, and it 
is indeed the intermediate regions that yield $\delta$ points 
with greater scatter in Fig.~3 and worse velocity reconstructions 
in Fig.~5. 

\begin{figure*}
\centering
\begin{picture}(360,360)
\includegraphics{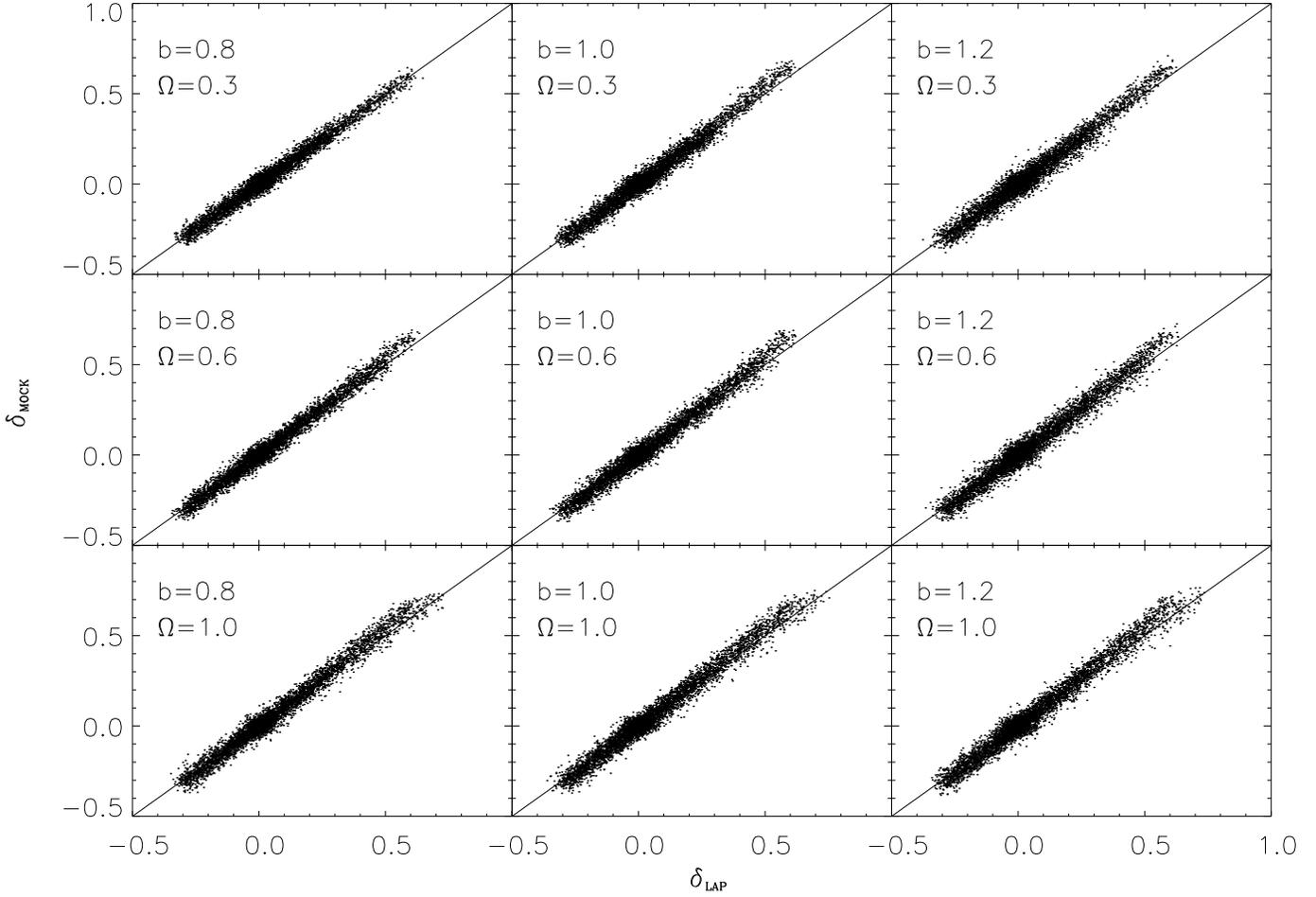}
\end{picture}
 \caption[]{Density field reconstructions of the nine datasets 
$d(b,\Omegam)$. The smoothed density contrast of the mock samples 
(vertical axis) is compared at each point within a selected 
spherical volume of the $128^3$ grid with the LAP-reconstructed 
densities (horizontal axis) over a sphere of radius $\sim 13,000$ \km.
The systematic errors are caused by the sparseness of the 
sampling.} 
\end{figure*}

\begin{figure*}
\centering
\begin{picture}(360,360)
\includegraphics{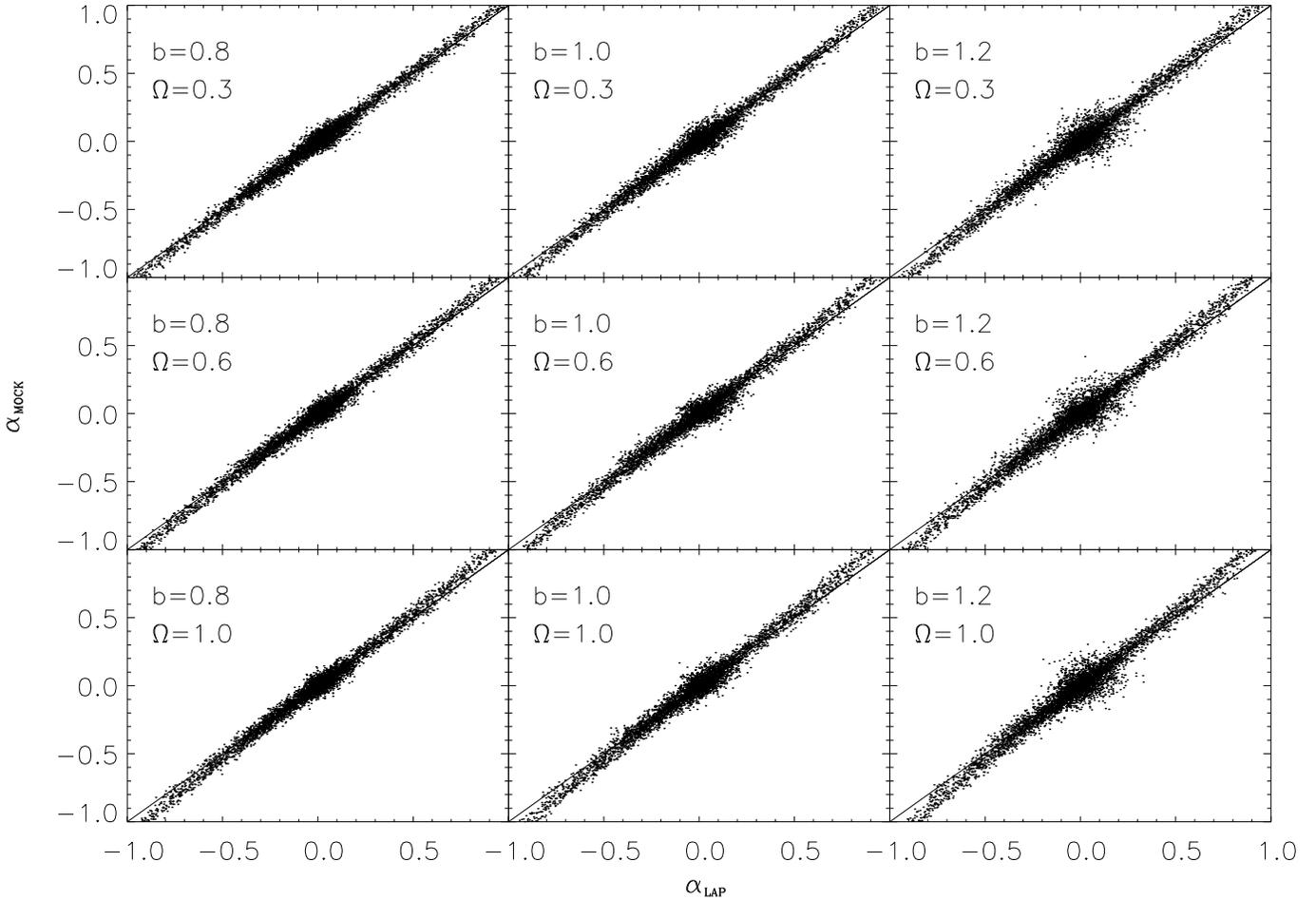}
\end{picture}
 \caption[]{Velocity potential field reconstructions of nine datasets 
$d(b,\Omegam)$. The velocity potential values are scaled to
$\alpha_{\rm max}$, so that they are dimensionless and consigned to 
the range $-1.0\lsim \alpha\lsim 1.0$. The smoothed 
velocity potential of the 
mock samples (vertical axis) is compared at each point within the 
same selected volume as in Fig.~3.} 
\end{figure*}

\begin{figure}
\centering
\begin{picture}(300,260)
\includegraphics{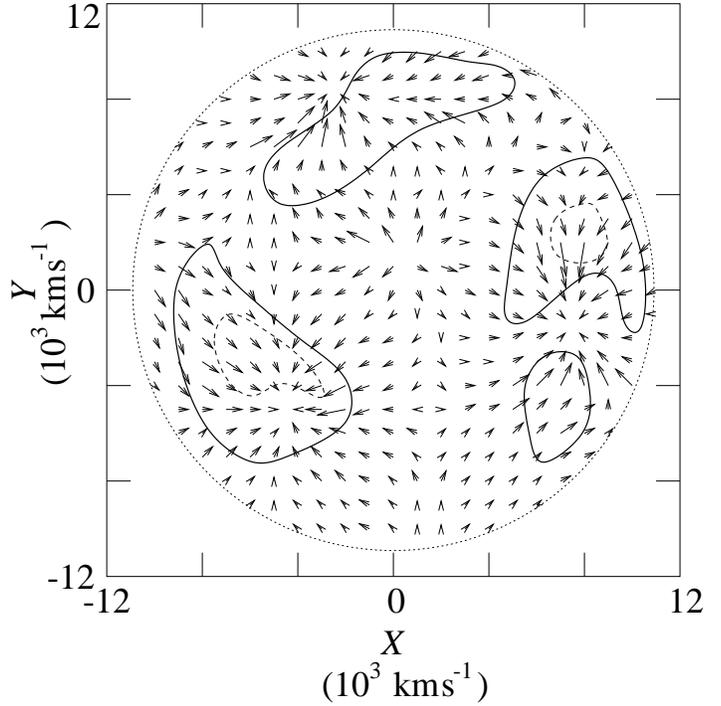}
\end{picture}
 \caption[]{$Z=0$ plane reconstruction of the velocity field within 
a selected subvolume $x_{\rm max}\lsim 12,000$ \km~for the mock
 sample $d(b=1.0,\Omegam=0.3)$. The solid lines enclose regions where 
the reconstruction entails an error 
$|{\b v}_{\rm mock}-{\b v}_{\rm LAP}|/v_{\rm mock}\gsim 0.10$ 
and within the broken lines this error is $\gsim 0.20$.}
\end{figure}

\section{Bias and $\Omegam$ from \iras}

We apply the LAP method to the \iras~ sample 
(Strauss \etal 1990,1992; Fisher \etal 1995a) in the same 
way as we have used it in the reconstruction of the mock catalogs 
in \S 3. \iras~ is not the largest existing near all-sky galaxy 
redshift catalogue, and it is now superseded by PSC$z$ 
(Canavezes \etal 1998) which contains $\sim 15,000$  
galaxies, so this application is simply an illustration on 
how the LAP method can be used to break the degeneracy 
in the estimates of $b$ and $\Omegam$. Other large redshift 
samples of partial coverage can also be looked at with the LAP method, 
e.g., Las Campanas and the forthcoming Anglo-Australian 2dF 
($\sim 250,000$ galaxies) and US Sloan Digital Sky Survey (SDSS) 
($\sim 10^6$ galaxies and 25\% coverage), 
with the caveat that boundary regions will be a source of 
propagating errors in the dynamical evolution. Even so, a large 
number of galaxies in a redshift survey of limited 
coverage can provide a good representation of the underlying 
density field, almost definitely outweighing the disadvantages 
of sampling a partial region of the sky, and it will be thus
predictably worthwhile to apply the LAP method to those surveys. 
The \iras~sample contains 5320 galaxies distributed over 87.6\% 
of the projected celestial sphere. The remaining unsampled 
2.4\% is an approximately disk-shaped region at a 
galactic latitude $|b|\lsim 5^{\circ}$. 

We adopt a Gaussian smoothing length of 1200 \km, and make no 
assumption regarding the power spectrum. The data $d_{\rm IRAS}$ 
are distributed within a spherical region of radius 
$x_{\rm max}\sim 15,000$ \km. We use the dataset in a similar 
fashion as the mock samples $d(b,\Omegam)$ in the previous section 
to derive the $x$-space fields $\delta,\alpha$. In \S 3 
we have established that $\sigma_{\delta},\sigma_{\alpha}$ 
are fairly insensitive to the values of $b,\Omegam$. One can thus 
set out to investigate the likelihood function $\lambda(b,\Omegam)$ 
as defined in (\ref{likelihood}) prior to determining the 
reconstructed fields. Evidently this is the simplest 
way to proceed for, unlike in \S 3, we do not have any clue about the 
real-space underlying fields (such as $\delta_{\rm mock}$,
$\alpha_{\rm mock}$ in \S 3) to compare them with the reconstructed 
fields.

\begin{figure}
\centering
\begin{picture}(300,200)
\includegraphics{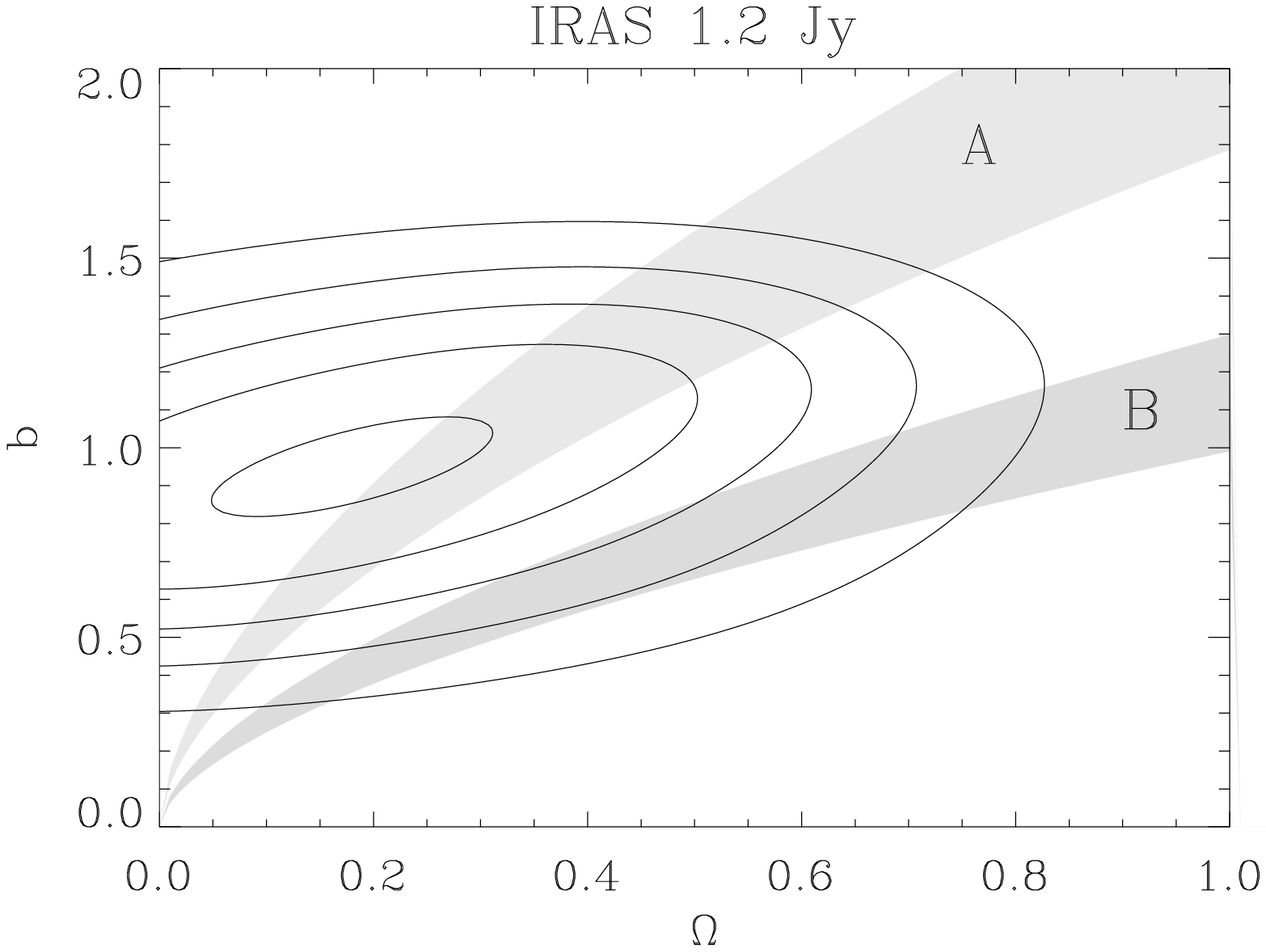}
\end{picture}
 \caption[]{Likelihood contours for \iras. 
The concentric contours represent a likelihood of 95\%, 75 \%, 
50\%, 25 \%, 10\% from the inner curve to the outer. The shaded 
region $A$ corresponds to the estimate of $\beta$ by Willick \etal 
(1997a) and the shaded region $B$ corresponds to an estimate of $\beta$ 
by Sigad. \etal (1998).}
\end{figure}

The likelihood contour plot is shown in Fig.~6. Clearly the largest 
values of the likelihood function are centered around $b\sim 1$ and 
small $\Omegam$. From the test of the LAP method in \S 3 with $n$-body 
simulations we already know that the likelihood function 
(\ref{likelihood}) underestimates both $b$ and $\Omegam$, as 
is apparent in all nine panels of Fig.~2. We accept this
trend is fairly inherent to the numerical application of the 
method and thus infer that the result presented in Fig.~6 is 
no different in this respect, 
and therefore the {\it real} values of the parameters are 
situated somewhat above their maxima in the likelihood function. 
From Fig.~2 one can quantify these errors to be of the order 
of $\Delta\Omegam \approx 0.12$, $\Delta b\approx 0.15$. Therefore, 
we infer that in Fig.~6 the likelihood maxima and the real values 
of the parameters are likely to be offset by a similar margin of 
error. At face value, Fig.~6 estimates that the most likely values 
of the parameters are $\Omegam\approx0.18$ and $b\approx 0.94$. 
If we offset these estimates by the errors derived from Fig.~2, 
then the likely ``real'' values of the parameters that we obtain 
for Fig.~6 are $\Omegam\approx 0.31$ and $b\approx 1.1$. As a matter 
of fact, these offset values are still within the region enclosed 
by the 95\% confidence contour. 

To put our results in perspective with previous analyses of \iras, we 
have overlaid on the contour plot of Fig.~6 two previous estimates of 
$\beta\equiv \Omegam^{0.6}/b$. An estimate by Willick \etal (1997a) 
yields $\beta_I=0.49\pm 0.07$ (shaded region $A$) and an estimate 
by Sigad \etal (1998) yields $\beta_I=0.89\pm 0.12$ (shaded region 
$B$). The estimate of Willick \etal (1997a) is clearly in better 
agreement with our results as the location of the offset maximum of 
the likelihood is contained within the shaded region $A$ that corresponds 
to the error margin of their estimate. The estimate given by shaded 
region $B$ is consistent with a scenario $b\approx 1$, $\Omegam\approx
1$, which in our analysis falls well outside the 10\% likelihood
contour. 

Fig.~7 shows a $z$-space comparison between the reconstructed 
fields and the dataset. The data on the horizontal axis, 
$\delta_{\rm LAP}^{s}$, is obtained from the reconstructed $x$-space 
fields $\delta,\alpha$ via (\ref{constraint}). The combination 
of both fields via the relationship $\delta_{\rm LAP}(\b x)  
\propto \delta_s (\b x+ \hat{x}\alpha_{\rm LAP}^{\prime})$ permits us to 
reconstruct $\delta_s$ which is our only possible point of comparison 
with $\delta_{IRAS}$, and this is shown in Fig.~7. The vertical 
axis shows the $z$-space data 
points of the smoothed \iras~sample. The data are plotted in a 
point-by-point comparison for all the grid points within the selected
subvolume. A solid line of slope 1.0 is plotted across the diagonal 
of the plot. The slope of the regression line is slightly over the 
diagonal line, at approximately 1.03. The corresponding rms due to 
random and numerical errors in the LAP reconstruction is 
$\sigma\approx 0.27$. The values of the parameters that have been 
used in the reconstruction are $b=1.0,\Omegam=0.3$.

\begin{figure}
\centering
\begin{picture}(300,200)
\includegraphics{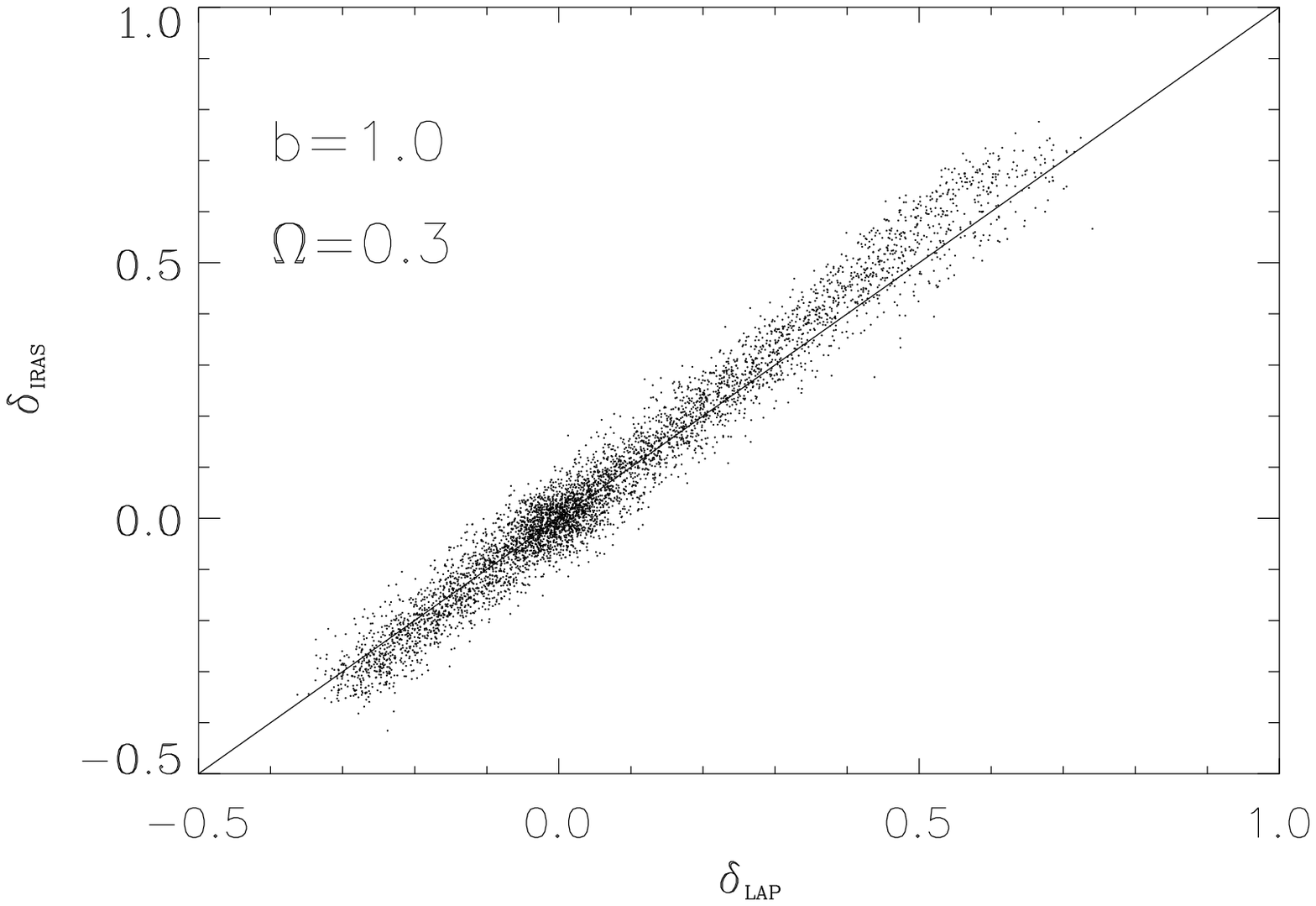}
\end{picture}
 \caption[]{Redshift-space density contrast in the LAP reconstruction 
versus the corresponding \iras~data for a Gaussian smoothing of 
1200 \km within a spherical region of radius 
$x_{\rm max}\sim$ 12,000 \km. The field $\delta_{\rm LAP}$ is 
evaluated at $b=1.0$, $\Omegam=0.3$.}
\end{figure}

\subsection{Velocity fields}

\begin{figure*}
\centering
\begin{picture}(550,550)
\begin{tabular}{cc}
\includegraphics{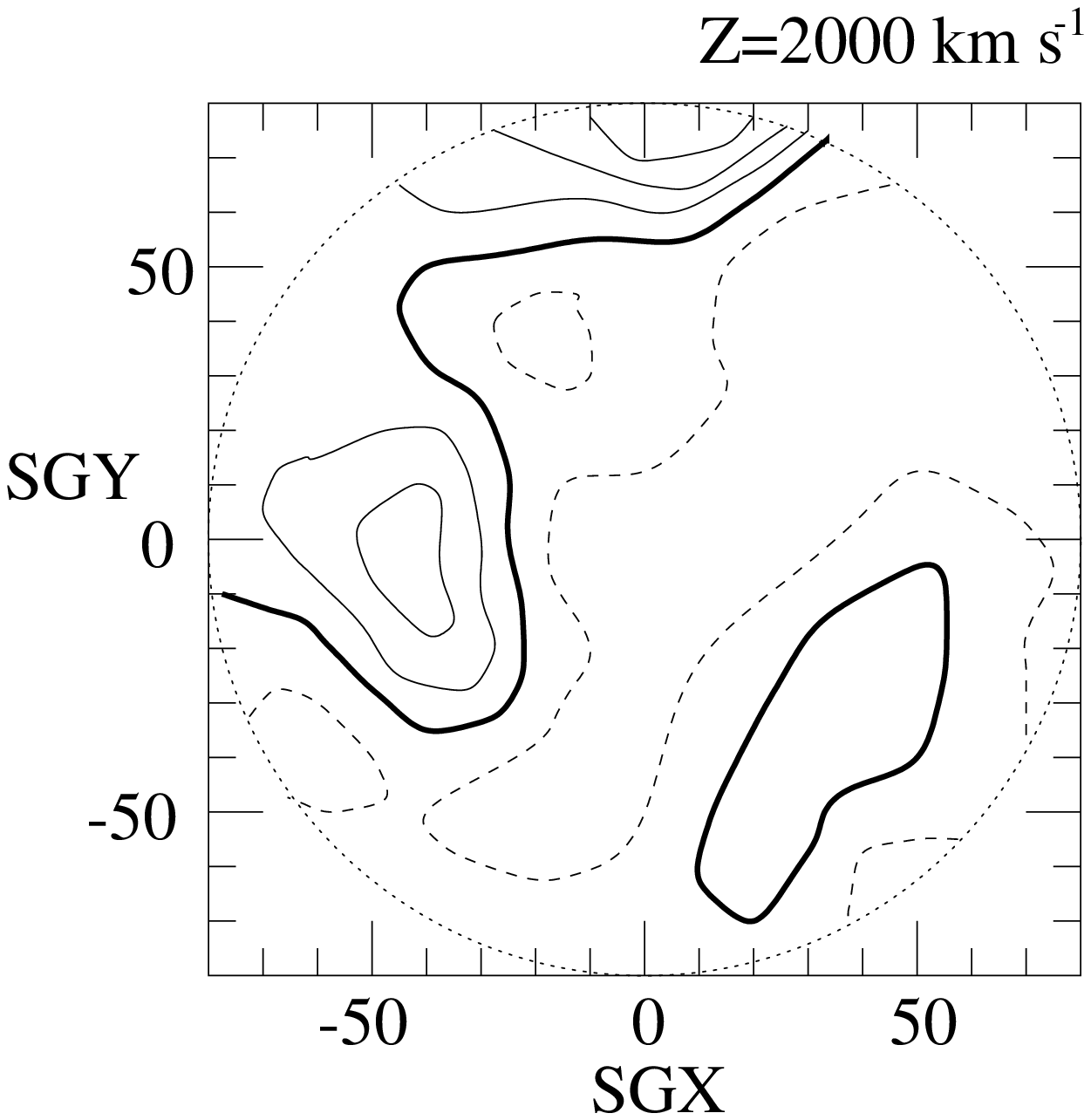}&
\includegraphics{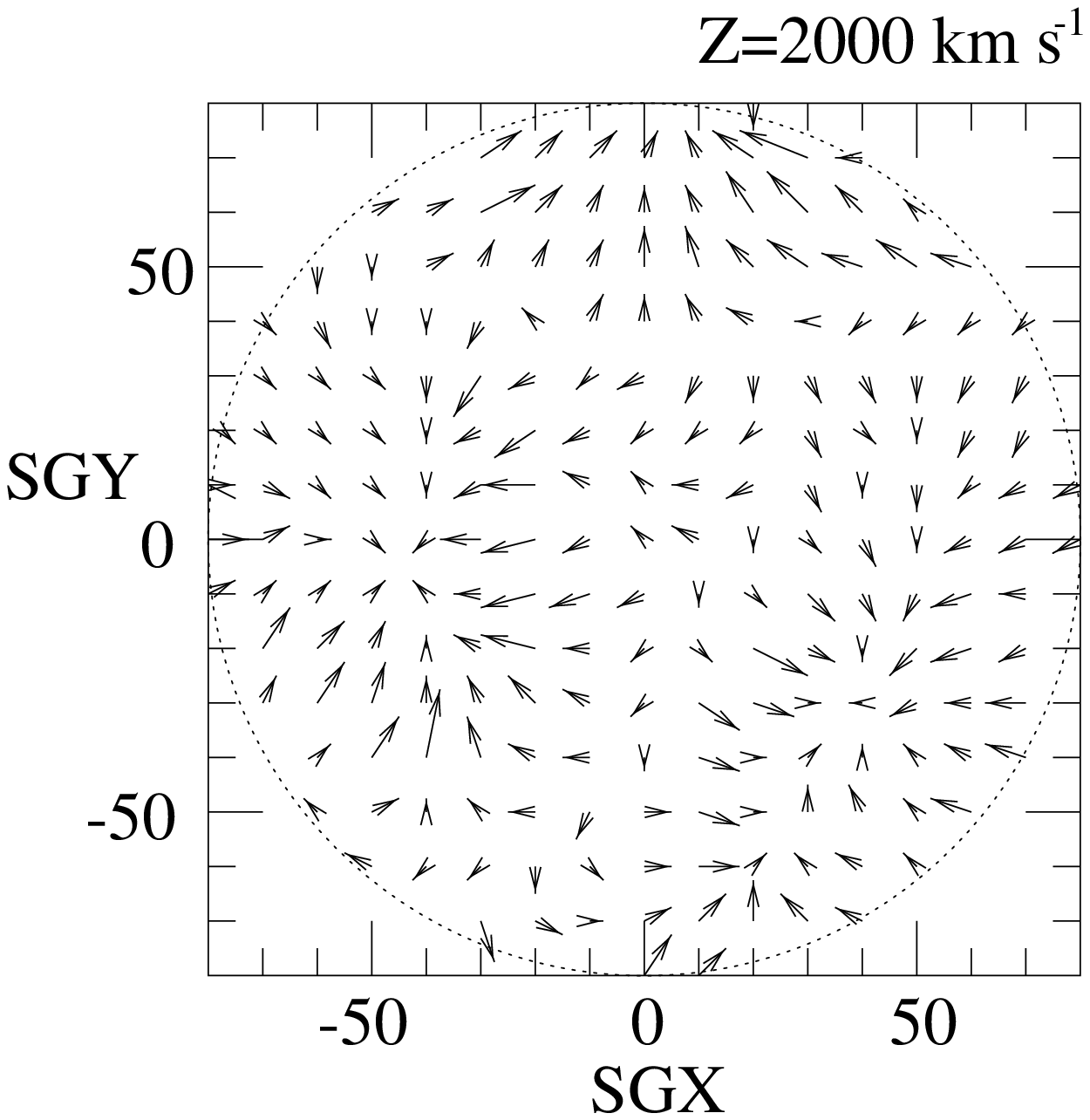} \\ 
\includegraphics{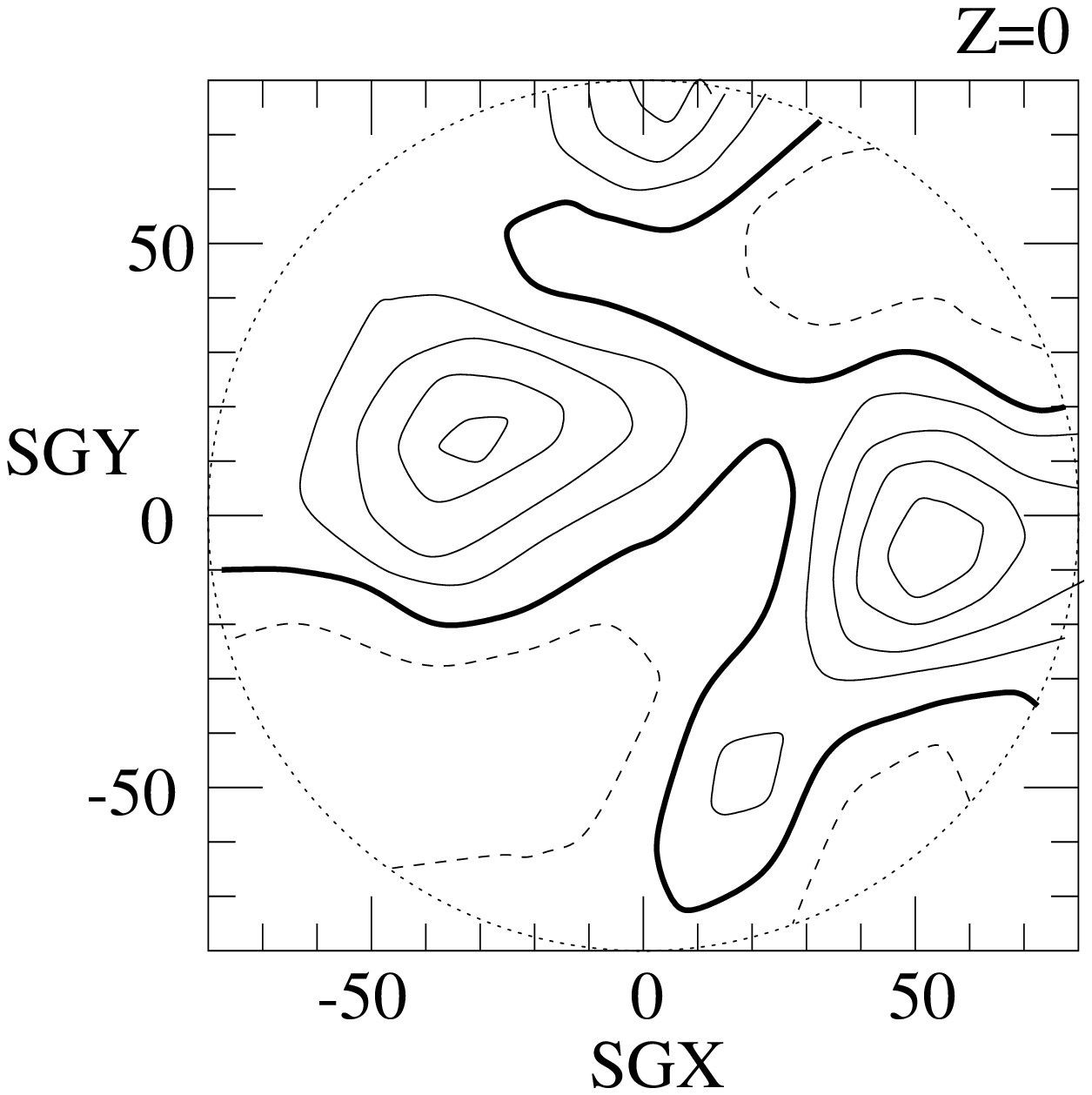} &
\includegraphics{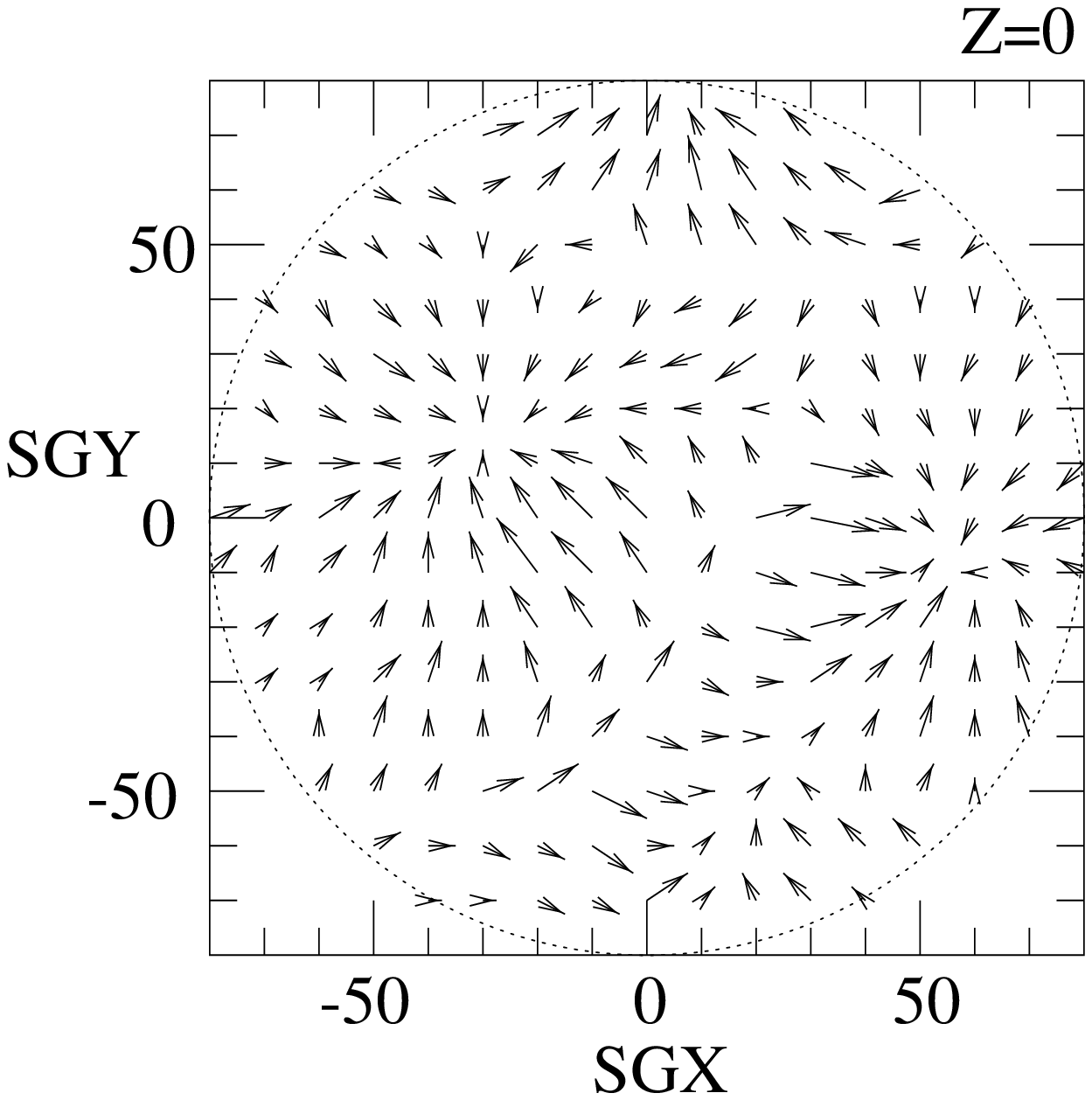} \\
\includegraphics{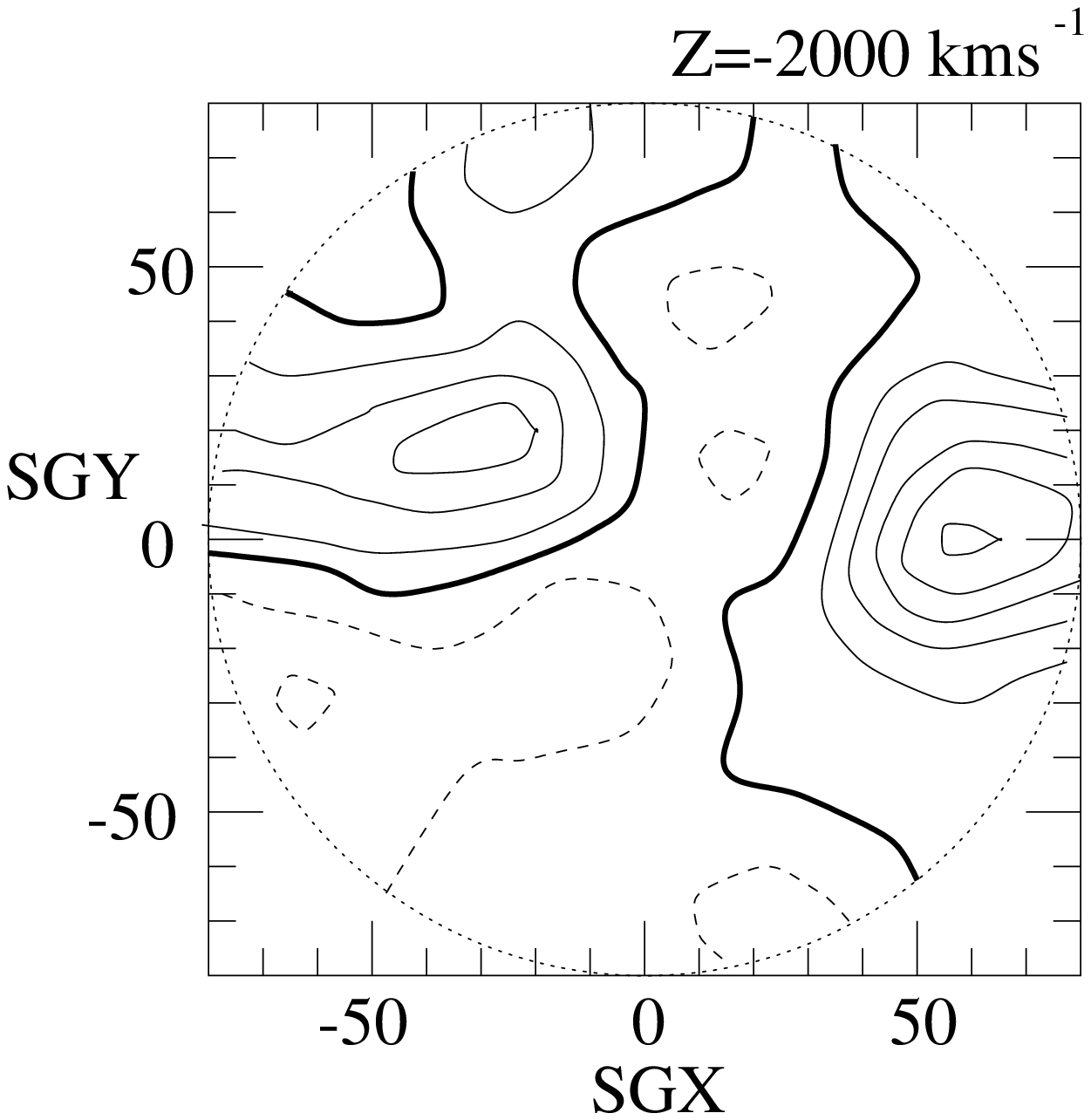} & 
\includegraphics{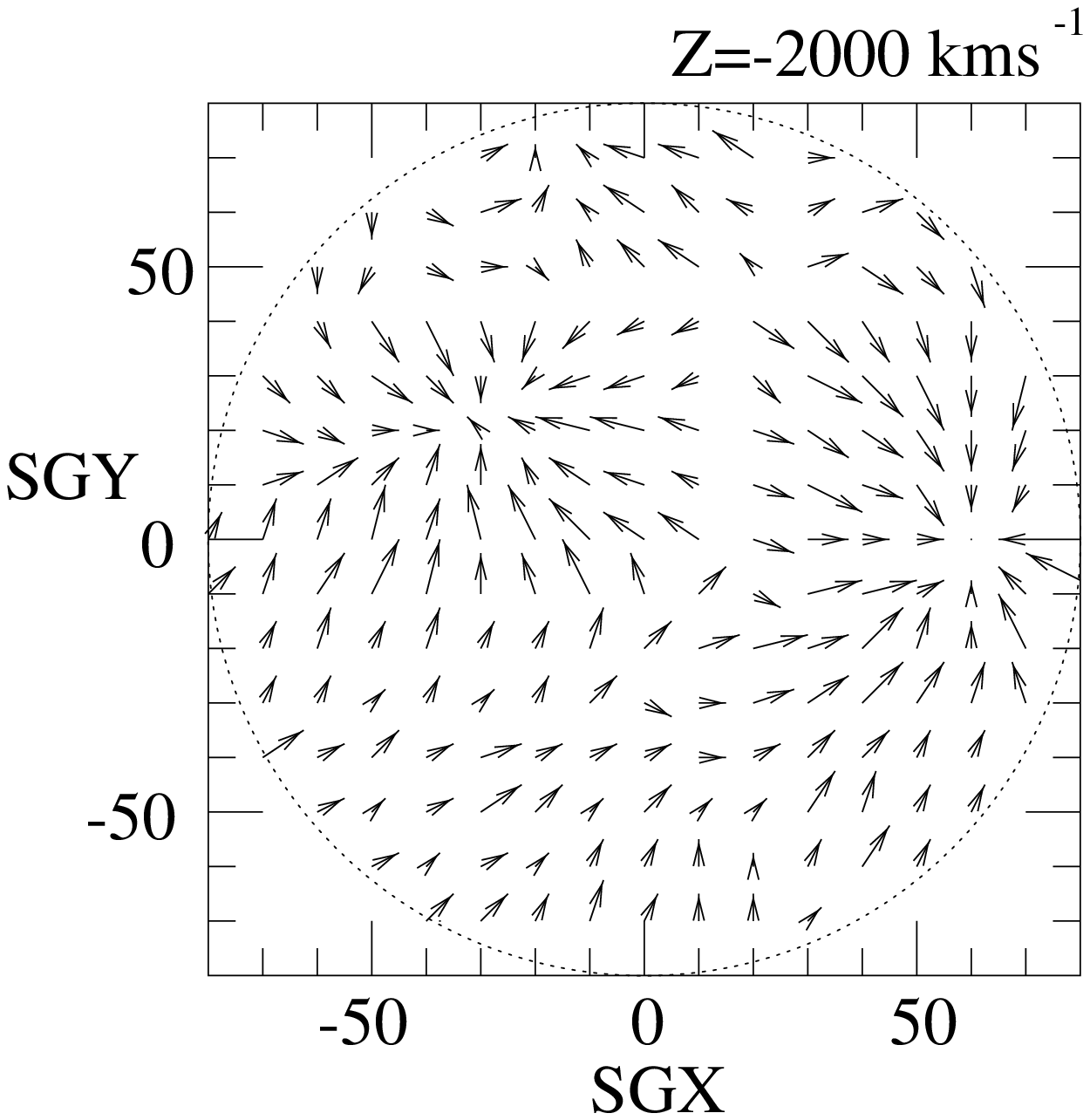} \\ 
\end{tabular}
\end{picture}
 \caption[]{$\delta_{\rm LAP}$ and $\b v_{\rm LAP}$ fields for \iras. 
$SGX$ and $SGY$ units are in 100 \km, spanning over a sphere 
of radius $\sim 8000$ \km. {\it Left:} from top to bottom panels, 
density contrast for a Gaussian smoothing of 1200 \km, 
for $Z= 2000,0,-2000$ \km. Thick solid line corresponds to 
$\delta=0$, continuous contours are $\delta>0$ and slashed 
contours are $\delta<0$; contour spacing is 0.2. {\it Right:} 
from top to bottom, reconstructed velocities at same values of $Z$.}
\end{figure*}

The resulting velocity field for the parameters 
of Fig.~7 is shown in Fig.~8. The six panels show the reconstructed 
\iras~fields $\delta_{\rm LAP}$ and $\b v_{\rm LAP}$ 
in supergalactic coordinates, for three different 
slices $Z=0,\pm 2000$ \km. The velocity panels on the right column 
correspond to the densities on the left, at the same value of $Z$. 
The velocity field follows the main features observed on the 
$\delta_{\rm LAP}$ field, with a general flow towards the overdense
regions and outflow from voids. The largest velocities are located in 
the intervening regions between overdense and underdense regions, e.g. in 
$Z=0$ (middle panels), large infall velocities are visible in the vicinity 
of the Comma supercluster (0,80,0), the Hydra-Centaurus 
(H-C) supercluster (-30,15,0), and Perseus-Pisces (P-P) (50,-5,0). 
In $Z=0$ the largest velocities are located at the lower right region 
of the H-C overdensity maximum, and also to the left of the P-P 
maximum. There is a velocity flow from the main void on the lower left  
of the figure, in the direction of Virgo, and it splits up to left 
and right, in manner of a ridge, to create an outflow in opposite 
directions, towards H-C and P-P. In the case of $Z=-2000$ \km 
(lower row), large velocities are also present around the steeper 
regions of the prominent overdensities, following a similar pattern 
as in $Z=0$, whereas the field shows more erratic features in 
$Z=2000$ \km (upper row), where the outflow from the main void 
(centre left) shows a general trend towards the main overdense 
features but is at the same time prone to local variations. 

The results presented in Figs.~6-8, can be optimized by using the 
Mark III velocity redshift survey to pin down $b$,$\Omegam$ 
more accurately. We shall pursue this and look for the optimal 
values of $b$,$\Omega_m$ by computing the LAP solutions that satisfy 
\be \label{vchi}
\delta \sum (\b v_{\rm LAP}-\b v_{\rm Mark III})^2=0, 
\ee 
where $\delta$ denotes a variation, not the density contrast. In 
practice, this is achieved as follows. One adds (\ref{vchi}) to 
the two already existing constraints of the LAP method 
(\ref{constraint1}),(\ref{constraint2}). Those are tackled in 
the manner summarized in \S 2.5. In actual terms, it's far 
more practical to deal with (\ref{vchi}) in terms of the 
velocity potential, so what we have done in the present analysis 
is in reality to compute $\alpha_{\rm Mark III}$ from the smoothed 
observed velocity field, and thus used (\ref{vchi}) in the manner 
of a second constraint on $\alpha$. 

The comparison with the $\b v_{\rm Mark III}$ data sets 
further constraints on  the likelihood contours of Fig.~6 as 
is shown below. Mark III contains approximately 3,400 galaxies, 
which are compiled from several sets of elliptical and SO galaxies 
(Willick \etal 1995,1996,1997a). The sample spans out to $\sim 6000$ 
\km, though in some directions it is irregularly sampled 
to $x_{\rm max} \sim 8000$ \km and $x_{\rm min}\sim 4000$ \km. 
The distances are inferred via forward Tully-Fisher and 
$D_n-\sigma$ distance indicators which may entail an error 
in the region 17-21\%. Mark III predicts a bulk flow 
$v_B \sim 194\pm 32$ \km towards the Shapley concentration 
(Zaroubi, Hoffman \& Dekel 1999)(for a low-resolution Gaussian 
smoothing $\sim 1200$ \km, within a sphere $r\sim 6000$ \km),  
in contrast to $v_B \sim 250-400$ \km~that is estimated in most 
other samples, including PSC$z$ (a compilation of $v_B$ 
estimates is summarized in Dekel 1999b). $\delta_{IRAS}$ 
and $\delta_{\rm Mark III}$ are consistent with mildly non-linear 
gravitational instability and linear bias (Sigad \etal 1998), 
though there are some differences, e.g. the Mark III sample 
appears to show a strong shear across the Hydra-Centaurus 
complex that is absent in \iras~(as indeed also in ORS). Recent 
papers have studied in detail the differences between the 
\iras~and Mark III velocity and density fields (Sigad \etal 1998; 
also Dekel \etal 1999 following an improved version of POTENT). 

\begin{figure}
\centering
\begin{picture}(300,200)
\includegraphics{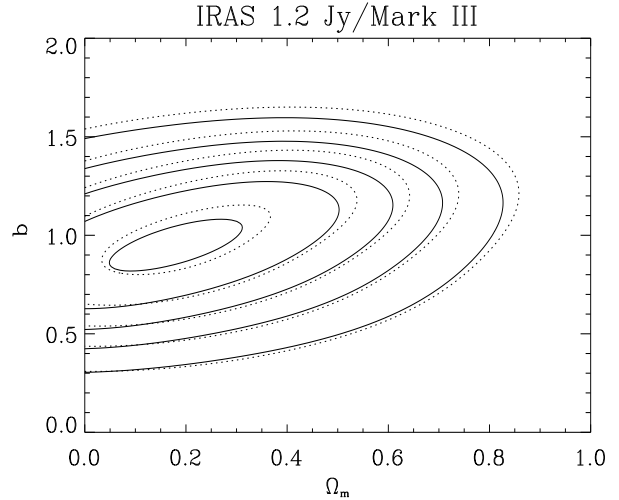}
\end{picture}
 \caption[]{Solid contours represent the likelihood 
for \iras~as in Fig.~6, and dotted contours represent 
the likelihood in the \iras/Mark III comparison 
following (\ref{vchi}). The relative likelihood of the 
concentric contours is as in Fig.~6 in both solid and dotted.}
\end{figure}

We consider the Mark III sample with a Gaussian smoothing 
length of 1200 \km. The data are carefully corrected for 
Malmquist biases (following the recipe set out in Sigad \etal 
(1998) for the preparation of the data), and the distances 
of 1,241 objects are modified as a result. The LAP method 
is solved for \iras~within spherical volume of radius 
$x_{\rm max}\sim 15,000$ \km, and the minimization fit 
with Mark III (\ref{vchi}) is done within a spherical 
subvolume of radius $\langle x\rangle \sim 6000$ \km. 
Therefore most of the volume of the LAP solutions remains free of 
the constraint (\ref{vchi}) and the fraction of the volume where 
$\b v_{\rm LAP}$ is least-squared to $\b v_{\rm Mark III}$ is only 0.064. 
Naturally such a small fraction forecasts an almost negligible impact 
in the fine-tuning of the parameters, unless the fields differred 
drastically to start with, which they do not. The $\b v_{\rm LAP}$ 
solution in the remainder of the volume is indirectly affected 
by this fit, and the variations in modulus $\Delta v_{\rm LAP}$ 
outside the comparison subvolume are $\lsim 12$\%. 

Fig.~9 shows the likelihood contours for ($b$,$\Omegam$) computed 
via the adjustment entailed in (\ref{vchi}). The solid contours 
are the purely \iras~prediction, as in Fig.~6, and the dotted 
contours are the result of the comparison with Mark III. The contours 
are ever so slightly shifted towards greater values of the 
parameters and, as expected, the effect is small. The shift 
towards larger $b$,$\Omegam$ is not in fact an altogether undesirable 
modification, as we have already discussed that the LAP solutions 
are found to be per se offset to smaller values than their ``real'' 
values. The important conclusion to be drawn from Fig.~9 is 
that the comparison with Mark III is entirely consistent with 
the predictions for $b$ and $\Omegam$ extracted from the \iras~sample 
alone.

\section{Conclusions}

The LAP method provides a practical means to break the 
degeneracy between $\Omegam$ and $b$ in galaxy redshift surveys.  
The method is employed in the manner of a nonlinear constraint 
on the redshift-space dataset and, although in formulation it 
comes across as algebraically cumbersome, it is of considerable 
simplicity and efficiency from the numerical point of view. The method is 
sound in that it does not require an a priori approximation of 
the map $\b x\to\b s$ to pin down the solution and it provides 
considerable freedom to ascribe relative importance to the data 
available, i.e. the initial and final endpoints, to which we wish 
to invariably assign greater weight than intermedate stages 
of which little or no data are available. 

The method can prove significant to measure $\Omegam$ in the latest 
largest samples, and extract the most accurate information prior 
to comparison with other datasets, such as the CMB radiation power 
spectrum and SN data. One important challenge for the future 
is to attain a better grasp of the concept of bias and this will 
be probably achieved via microlensing data and $n$-body simulations 
of the formation of galaxies and clusters from primordial
fluctuations, rather than from galaxy redshift surveys. Once a 
model of bias is adopted on a sound footing, then clearly the 
LAP model is impeccable in producing an estimate of $\Omegam$. 
In the simple linear bias model we have employed we have totally 
relegated any consideration of scale-dependence in $b$. This 
is a point I have deliberately omitted for simplicity. Thus, 
the estimates computed in this paper ought to be regarded 
qualitatively as weighted averages of the ``real'' $b$ over 
different scales, if indeed scale-dependent bias models are 
to be believed. 

In this paper, we have employed the likelihood function 
(\ref{likelihood}) to investigate the values of $b$,$\Omegam$. 
Clearly this is not a unique choice. However, our choice is 
guided by the argument of relative convergence of the solutions, 
which is justifiably a reasonable criterion to get close to 
the ``real'' solutions. In view of the performance of the $\lambda$ 
function in the reconstruction of the mock samples, this 
choice does not appear to be totally off the mark. A potential 
reason for concern could be the offset observed between the maxima 
of the likelihood functions and the real values of the parameters 
in the $n$-body simulations. However the recurrence of this offset 
in a predictable manner lends strength to the argument that 
it arises as a numerical fault that is easy to account for
systematically in the analysis of the datasets. The reconstructions 
of the fields are, on the other hand, of considerable accuracy 
and no numerical defficiency or hindrance is observed. 
The application of the method to \iras~predicts the
parameters to be fairly accurately located in the immediate 
neighbourhood of the maxima $\Omegam\approx 0.3$ and 
$b\approx 1.1$, which is found to be most compatible with 
the estimate of $\beta$ given by Willick \etal (1997a). 
In a flat universe such predicted values are perfectly consistent 
with a non-vanishing cosmological constant or a quintessence 
scalar field component. The likelihood examined in this way is only very 
slightly modified when the velocities predicted via the LAP method 
are finely-tuned with data from the Mark III sample. The shift of 
the predicted values is towards slightly greater values of the
parameters but it remains comfortably consistent with the results 
obtained from \iras.

\section*{Acknowledgments}

The numerical analysis has been performed in the {\it Starlink}  
facilities at Queen Mary \& Westfield College (London), and 
thanks are due to a number of people for supplying 
and lending a helping hand with technicalities regarding the data. 
The author thanks an anonymous referee for insightful comments 
that have greatly improved the readability of the manuscript.  
This research has been funded in part at EHU by research grant 
UPV172.310-G02/99.

\appendix

\section[]{Orbit-crossing in redshift space}
We shall prove the boundary condition (\ref{constraint}). 
The number-counts of galaxies $n$ in $x$-space and $z$-space satisfy, 
by conservation of the number of galaxies:
\be \label{nsnx} 
\d n(\b s)=\sum_{\rm streams}\d n(\b x_i),  
\ee
for all streams at the same redshift, 
$\b s=\b x_i+\hat{x}(\hat{x}\cdot\nabla_{x}\alpha_i)$. 
In our analysis we shall only consider single-valued solutions, 
and therefore there is just one stream only in (\ref{nsnx}), 
i.e. $\d n(\b s)=\d n(\b x)$. Hence 
\be \label{masscons}
\rho_s(\b s)\,\d{\sl \Omega}\equiv {\d n(\b s)\over\d s}=
x^2(1+b\delta)\gal {\d x\over\d s}\d{\sl \Omega},
\ee
where $n(\b s)$ is the galaxy number-count, $\d{\sl \Omega}$ a solid angle 
element and the $x$-space selected volume of the sample is 
$V\sim {4\over 3}\pi x_{\rm max}^3$, and 
\be
s\equiv \hat{x}\cdot\b s=x+\alpha'.
\ee
Therefore
\be
{\d x\over\d s}={1\over 1+\alpha''},
\ee
and substituting this in (\ref{masscons}), we get 
\be \label{qed}
\rho_s =x^2 \gal \biggl({1+b\delta\over 1+\alpha''}\biggr).
\ee
In the case of multistreams, the RHS of (\ref{qed}) is integrated 
over all streams, bearing in mind that {\it turn-around} regions, 
which occur at $\delta\gg 1$ and for which $\d s/\d x=0$, are 
excluded from the sum. An example of such a region in shown in Fig.~A1. 
An initial saddle-point $\d s/\d x=0$ on the $s(x)$ curve starts the 
creation of a turn-around region. At the stage shown in Fig.~A1, 
both points $A$ and $B$ satisfy this condition and obviously 
they departed from an initial saddle-point $A=B$. The region 
spanning between $A$ and $B$ is three-valued (each redshift 
in the interval $z_B < z < z_A$ corresponds to three $x$ positions), 
whereas $z_A$ and $z_B$ are bivalued. To make such 
an scenario tractable, we need to replace $s(x)$ over the interval 
$z_B<z<z_A$ by a monotonic curve that matches the existing curve at 
$z_B$ and $z_A$ and its first derivative. This is obviously tantamount to 
applying a larger smoothing length than the existing one to erase the 
overdense region that is the cause of the turn-around.   

\begin{figure}
\centering
\begin{picture}(240,240)
\includegraphics{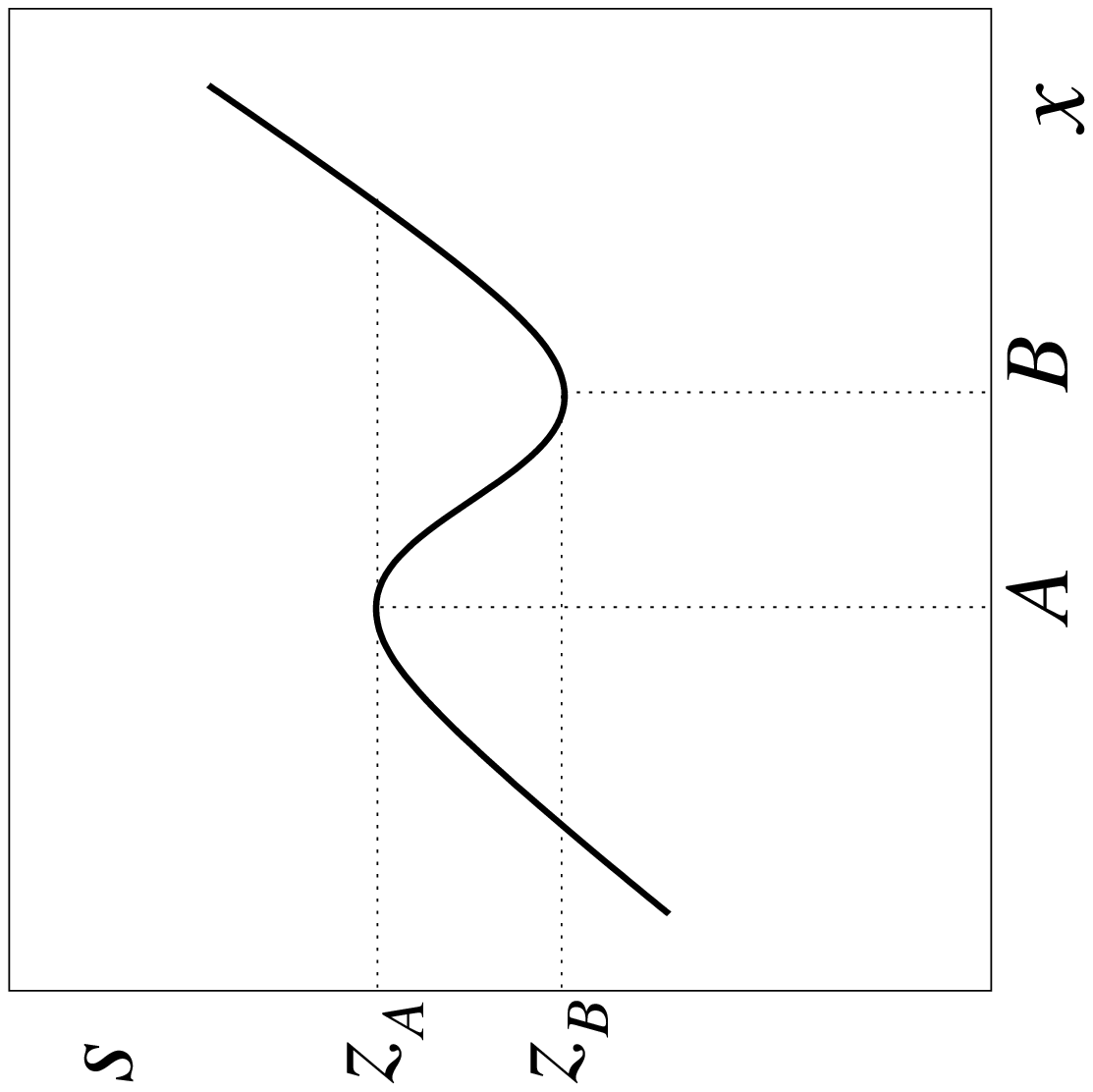}
\end{picture}
 \caption[]{Illustration of a turn-around region.}
\end{figure}

\section[]{Evaluation of radial derivatives}
The radial derivative of the velocity potential coefficients 
(\ref{zalphan}) can be written as 
\be
{\d\over\d x}\alpha_n =
\sum_{rlm}\alphap_{rlm}^{(n)}j_l(k_rx) Y_{lm},
\ee
where, using the equality 
\be \label{jprime}
{\d\over\d u}j_l(u)=(2l+1)^{-1}
\Big[ lj_{l-1}(u)-(l+1)j_{l+1}(u)\Big],
\ee
we have  
\be \label{alphap}
\alphap_{rlm}^{(n)}=k_r\biggl[{(l+1)\over
(2l+3)}\alpha_{r(l+1)m}^{(n)} 
-{l\over(2l-1)}\alpha_{r(l-1)m}^{(n)}\biggr].
\ee
Similarly
\[
\alphapp_{rlm}^{(n)}= k_r^2\biggl\{{(l+1)
\over(2l+3)}{(l+2)\over (2l+5)} \alpha_{r(l+2)m}^{(n)}
\]
\be
-\biggl[{(l+1)^2\over(2l+3)(2l+1)}+{l^2\over(2l-1)(2l+1)}\biggr]
\alpha_{rlm}^{(n)}
\ee
\[
+{l\over (2l-1)}{(l-1)\over(2l-3)}\alpha_{r(l-2)m}^{(n)}\biggr\}.
\]
On the other hand, the coefficients $\b J_{lm}^{(n)}$ given in 
(\ref{v_spher}) are 
\[
\b J_{lm}^{(n)}=\sum_r\biggl[{\alpha(l,m+1)\over 2}\alpha_{rl(m+1)}^{(n)}
\,(i\hat{x}_1-\hat{x}_2)
\]
\be \label{jlmn}
+{\beta(l,m-1)\over 2}\alpha_{rl(m-1)}^{(n)}
\,(i\hat{x}_1+\hat{x}_2)+im\,\alpha_{rlm}^{(n)}\,\hat{x}_3\biggr],
\ee
where 
\be
\alpha(l,m)=\Big[l(l+1)-m(m-1)\Big]^{1/2},
\ee
\be
\beta(l,m)=\Big[l(l+1)-m(m+1)\Big]^{1/2}. 
\ee

\section[]{Chebyshev polynomials}

The Chebyshev polynomials are defined  
$T_n(\cos\theta)\equiv \cos (n\theta)$ (following the normalization 
of Abramowitz \& Stegun 1972). We define the angle brackets 
$\langle,\rangle$ according to the orthogonality 
properties of $T_n$ (e.g. Courant \& Hilbert 1989): 
\be
\langle u\rangle \equiv \int_{-1}^{1}\d t w(t) u(t),
\ee
where $w(t)=(1-t^2)^{-1/2}$ is a weight function and therefore  
\be \label{TT}
\langle T_n T_m\rangle =\delta_{nm}{\pi\over 2},
\ee
for $n\neq 0$ and $\langle T_0^2\rangle =\pi$. In 
(\ref{zshortI})(\ref{zshortII}) we encounter two types of angle 
brackets to evaluate (other than (\ref{TT}): 
$\langle T_n\dot{T}_m\rangle$ and $\langle T_n T_m T_r\rangle$ 
(we have deliberately omitted $\langle \Omegam T_n T_m\rangle$, by 
approximating $\Omegam$ by a constant, and 
ditto for $H$. The second type 
of product is trivially transformed into (\ref{TT}) via 
\be 
2 T_n T_m = T_{n+m}+T_{n-m}
\ee 
for $n\geq m$, and the first requires a little numerical 
manipulation using the relation
\be 
(1-t^2)\dot{T}_n = -nt\, T_n + T_{n-1}. 
\ee 

\section[]{Orthogonality relations}

The orthogonality relations for the spherical harmonics and 
the Bessel functions are respectively 
\be  \label{orthoI}
\int_0^{2\pi}\!\!\d\varphi\int_0^{\pi}\!\!\d(\cos\theta)
Y_{lm}Y_{l'm'}=\delta_{ll'}\delta_{mm'},
\ee
\be   \label{orthoII}
\int_0^1\!\!\d x x^2 j_l(k_rx)j_l(k_sx)
={1\over2k_rk_s} \Big[j_l(k_rx)+xj_l'(k_rx)\Big]^2\delta_{rs}.
\ee

\bsp

\label{lastpage}

\end{document}